\documentclass{aa}  
\usepackage{blindtext}

\usepackage{graphicx}
\usepackage{txfonts}

\usepackage[hidelinks,colorlinks=true,linkcolor=blue,citecolor=blue]{hyperref}

\newcommand{\mstar}{M$_{*}$}
\newcommand{\Msun}{\rm M_{\sun}}

\newcommand{\ropt}{R$_{\rm opt}$}
\newcommand{\cielo}{{\sc CIELO}}

\newcommand{\sMZR}{MZ$_{\mathrm{*}}$R}
\newcommand{\facc}{$f_{\rm acc}$}
\newcommand{\fout}{$f_{\rm ex-situ}$}
\newcommand{\fmig}{$f_{\rm DB}$}

\begin{document} 

   \title{The mass-metallicity relation of bulges}

   \author{Ignacio Mu\~noz-Escobar 
          \inst{1}\fnmsep\inst{2}\thanks{E-mail: imuoz@uc.cl}
          \and Patricia B. Tissera\inst{1}\fnmsep\inst{2} \and 
          Jenny Gonzalez-Jara\inst{1}\fnmsep\inst{2}  
          \and 
          Emanuel Sillero\inst{1}\fnmsep\inst{2}  
          Valentina P. Miranda\inst{1}\fnmsep\inst{2} \and 
          Susana Pedrosa\inst{3}  \and 
          Lucas Bignone\inst{3}}
        \institute{Instituto de Astrof\'isica, Pontificia Universidad Cat\'olica de Chile. Av. Vicu\~na Mackenna 4860, Santiago, Chile.
         \and
Centro de Astro-Ingenier\'ia, Pontificia Universidad Cat\'olica de Chile. Av. Vicuña Mackenna 4860, Santiago, Chile.
\and
Instituto de Astronom\'ia y F\'isica del Espacio, CONICET-UBA, Casilla de Correos 67, Suc. 28, 1428 Buenos Aires, Argentina}

\date{Received September 7, 2025; accepted November 09, 2025}
 
  \abstract{\textit{Context.} Bulges, located at the central regions of galaxies, are complex structures, expected to be shaped by the physical processes involved in the assembly history of their host galaxy, such as gravitational collapse, mergers, interactions, and bars.  As a consequence a variety of bulges with distinct morphology and chemistry could be produced.
  
\textit{Aim.} We aim at  exploring the existence of a stellar mass-metallicity relation, \sMZR, of bulges,  and analyze the possible imprint of characteristics features by  accretion and migration of stars, which could store information on their assembly histories.

\textit{Methods.} We use 44 central galaxies from the \cielo~cosmological simulations. Their stellar masses are within the range of $\sim [10^{7.6}, 10^{10.6}]$ M$_\odot$. We decomposed the galaxy into bulge and disk using the circularity and binding energies. We track the stellar populations in bulges back in time to their birth location, classifying them  as bulge-born in-situ, and disk-born stars and accreted.

\textit{Results.} We find that most of the stars in our bulges are formed in-situ, but 33\% of our bulges show a non-negligible contribution of stellar accretion from satellites, which could add to about 35\% of the population. The accreted material is generally contributed by two or three satellites at most. In some bulges, we  also find up to a 32\% of stars that migrated from the disk due to secular evolution, with a median of 10\%. 
Regardless of the formation histories, we found a clear \sMZR~for bulges, which is more enriched by about 0.4 dex than the corresponding relation of the disk components, and about  0.15 dex more enriched than the galaxy \sMZR.
We find evidence that the dispersion in the bulge \sMZR~is influenced by both stellar accretion from satellites and migration from the disk, such that, at a fixed bulge mass, bulges with higher fraction of accreted and migrated stars tend to be less metal-rich. Therefore we find a \sMZR~for bulges, which is consistent with an increase of metallicity with increasing mass, while its dispersion stores information on the contribution from different formation channels.
  }
               
   \titlerunning{Stellar mass-metallicity relation of bulges}
   \maketitle
%

\section{Introduction.}
The observed diversity of galaxy morphology could be linked to their different formation paths. In the local Universe, galaxies are traditionally classified using the Hubble sequence, which arranges them along a morphology, from spheroidal to disk-dominated types \citep{1926:Hubble} .
The central region of  disk galaxies, so-called the bulge, has distinct kinematics and chemical abundances.  
Early studies classified the bulge as analogous to an elliptical galaxy \citep{1962:Eggen,1999:Noguchi}. However, this view has been challenged by more recent studies, which show that bulges are more complex stellar structures and often exhibit significant rotational support. Bulges are commonly classified in two broad categories: classical bulges, which present a dispersion-dominated stellar populations, and pseudo-bulges, which correspond to a composite  of a classical bulge, a disk-like component, and sometimes a bar \citep[e.g.,][]{2010:Fisher,2016:Kormendy,2020:Breda}. 
The Milky Way is a clear example of a galaxy with a complex bulge where the three mentioned components coexist \citep{2017:Zocalli, rojas2019,2021:Queiroz}

The formation of the dispersion-dominated component of bulges has been attributed to fast and violent processes, such as the gravitational collapse of a proto-galaxy \citep{1999:Avila-Reese}, or major mergers that triggered strong starbursts \citep{2008:Hopkins, 2014:Abreu}. 
Their origin appears to be distinct: numerical simulations report the formation of pseudo-bulges with varying degrees of rotation \citep{2011:Guedes, 2019:Gargiulo}, while secular processes have been found to play a positive role in their build-up \citep[e.g.,][]{2004:Kormendy, 2011:Eliche, 2019:Rosito, 2022:Gargiulo}. Together, these results suggest that bulge formation proceeds through multiple evolutionary pathways, ranging from rapid and violent events to slower, secular processes shaping the central regions of galaxies.
Additionally, the formation of clumps in early gas-rich disks, which later migrate and fall into the central regions of galaxies, has been reported as another channel for bulge growth \citep{2011:Bournaud,2013:Perez,2023:Debattista}. Since different mechanisms may operate in bulge formation, with potentially distinct efficiencies as a function of redshift \citep[][]{2023:Debattista}, their chemical signatures could provide a valuable diagnostic to disentangle the relative contributions of these processes.

The chemical evolution of galaxies is closely linked to the physical processes that drive galaxy assembly \citep[e.g.,][]{1980:Tinsley, 1997:Pagel, 1997:Chiappini, 2019:Maiolino}. One of the most fundamental relations in this context is the mass–metallicity relation (MZR), which connects the galaxy stellar mass to the metallicity of its star-forming gas \citep{lequeux1979, skillman1989,zaritsky1994,2004:Tremonti}. This correlation has been well established for decades, both in the local universe \citep[e.g.,][]{2004:Tremonti, 2014:Zahid} and at high redshift \citep[e.g.,][]{2024:Lewis, 2024:Cheng}. The prevalence of this relation has been studied up to redshift $z \sim 7$, and beyond with the contribution of the James Webb Space Telescope \citep{2024:Stanton}. It is commonly interpreted as reflecting the ability of higher-mass galaxies to retain metals more efficiently, thereby reaching higher metallicities.  Numerical simulations are able to reproduce the MZR at least in the low redshift universe \citep[e.g.,][]{2017:DeRossi, 2022:Zenocratti}. Analogously to the gas, the stellar mass-metallicity relation is likewise in place  \citep{2005:Gallazzi, 2013:Kirby}. This relation has a similar shape to the MZR, but it reflects the cumulative evolutionary history of the interstellar medium (ISM), since stars inherit the chemical abundances of the gas from which they form. The stellar metallicity  trace the integrated history of the stellar populations in a galaxy. Traditionally, stellar metallicity has been derived by fitting spectral energy distributions of composite stellar populations to the integrated spectra of galaxies \citep{2013:Conroy}. However, in recent years, with the use of integral field units (IFUs), it has become possible to determine the two-dimensional distribution of stellar population properties, giving rise to the so-called resolved relations MZR \citep{gonzalezdelgado2014,campas2021, 2023:Baker}. 
 Recently, \citet{2024:Jegatheesan} analyzed a large sample of galaxies from the MaNGA survey and found that galaxies with different morphologies follow distinct assembly histories, with their bulge and disk components displaying a wide diversity of metallicities and ages.

In this paper, we aim at studying the \sMZR~of bulges for the first time to our knowledge. We analyze the relation between the scatter of the \sMZR~and the assembly history of the bulges.  While most of the stellar populations in the bulges are expected to form in-situ \citep{2019:Gargiulo}, a fraction could form in the disk and then migrate \citep{2023:Debattista} or could even be accreted \citep{romano2023}. If bulges are composed by stars formed  from different channels, they might leave an imprint in the bulge \sMZR.

This paper is organized as follows. Section 2 describes the simulations and the galaxy sample. In Section 3, we analyze the bulge \sMZR\ and the contribution of stellar populations originating from different channels. Section 4 explores the various formation scenarios. Finally, Section 5 presents our conclusions and summarizes the main findings.

\section{The \cielo~galaxies}

As mentioned above, cosmological simulations are powerful tools to study the aforementioned diversity of processes that could take place in the formation of bulges \citep[e.g.,][]{tissera2006,2022:Gargiulo, 2017:Fragkoudi}. Hence,
in this work, we use the cosmological simulations from the Chemo-dynamIcal propertiEs of gaLaxies and the cOsmic web, the \cielo~Project \citep[][]{2025:Tissera}. 
These set of simulations has been previously used to study the evolution of galaxies in the Local Group  \citep{2022:Rodriguez}, the impact of baryons on the dark matter shape \citep{2023:Cataldi}, the channel of formation of stellar halos \citep{2025:Gonzalez}, the shape of the metallicity gradients \citep{2025:Tapia} and the possible contribution of Primordial Black Holes to the dark matter component \citep{2024:Casanueva}.  We note that the \cielo~galaxies have been already shown to determine a stellar and star-forming gas MZR in global agreement with observations \citep{2025:Tissera}.

Here we provide a description on the \cielo~simulations and the galaxy sample selected for this work.

\subsection{The simulations}
The  \cielo~project encompass cosmological hydrodynamical simulations of zoom-in regions consistent with a $\Lambda$-CDM cosmological scenario 
with $\Omega_{0} = 0.317$, $\Omega_{\Lambda} = 0.6825$, $\Omega_{B} = 0.049$ and $h = 0.6711$ \citep{2014:Planck}. The \cielo~ project includes zoom-in regions selected from a 50 Mpch$^{-1}$ side-box volume to cover a diversity of environments while avoiding massive groups or clusters, as well as two Local Group Analogs extracted from a 100  Mpc h$^{-1}$ side-box volume.  The target halos were selected to have virial masses in the range $[10^{11} - 10^{13}]\Msun$ and the surrounding region of about 1 Mpc radius was resampled at the same level of resolution. The initial conditions are explained in detail in \citet{2025:Tissera}. Simulations with two  numerical resolution levels are used,  with  dark matter particles of m$_{\mathrm{dm}} = 2 \times 10^{5}\Msun$ and  m$_{\mathrm{dm}} = 1.91 \times10^{6}\Msun$ for the high and intermediate resolution level. The initial gas masses are m$_{\mathrm{gas}} = 3.12 \times 10^{4}\Msun$ and  m$_{\mathrm{gas}} = 2.98 \times10^{5}\Msun$, respectively. For the high resolution level the gravitational softening length is $\epsilon_\mathrm{dm} = 500$ pc for dark matter, and  $\epsilon_\mathrm{gas} = 250$ pc for gas and stars. The intermediate resolution one has values of $\epsilon_\mathrm{dm} = 800$ pc and  $\epsilon_\mathrm{gas} = 400$ pc. 

Details of the simulations  can be found in \citet{2025:Tissera}. Briefly, the simulations were performed by using a version of {\sc GADGET-3} \citep{2003:Springel, 2005:Springel} that includes  metal-dependent radiative cooling and   a multiphase treatment for the ISM
\citep[][]{2006:Scannapieco}. A chemical enrichment and energy feedback  by Type Ia (SNIa) and Type II (SNII) supernovae are also included \citep{2001:Mosconi, 2005:Scannapieco}, following 13 individual isotopes. 
SNII are assumed to originate from stars more massive than 8 ${\rm M_\odot}$, with nucleosynthesis products from the metallicity-dependent yields of \citet{1995:WW} and lifetimes from the mass–metallicity–dependent prescriptions of \citet{1996:Raiteri}. For SNIa,  the W7 model \citep{1999:Iwamoto} was adopted, assuming progenitor lifetimes randomly distributed in the range 0.7–1.1 Gyr, see \citet{2015:Jimenez} for a discussion on this. The number of SNIa events is estimated from an observationally motivated SN II-to-SN Ia rate ratio \citep{2001:Mosconi}. The adopted initial mass function is consistent with \citet{2023:Chabrier}.

The virial halos were identified by using the Friend-of-Friends algorithm \citep[FoF,][]{1985:Davis}   and the SUBFIND code \citep{2001:Springel,2009:Dolag} was applied  to select subhalos within the virial radius. The merger trees were built by using the AMIGA package\citep{2009:Knollmann} as explained by \citet{2025:Tissera}. We use the \cielo~data base prepared by \citet{2025:Gonzalez}.

\begin{figure}
   \centering
        \includegraphics[width = 0.5
        \textwidth]{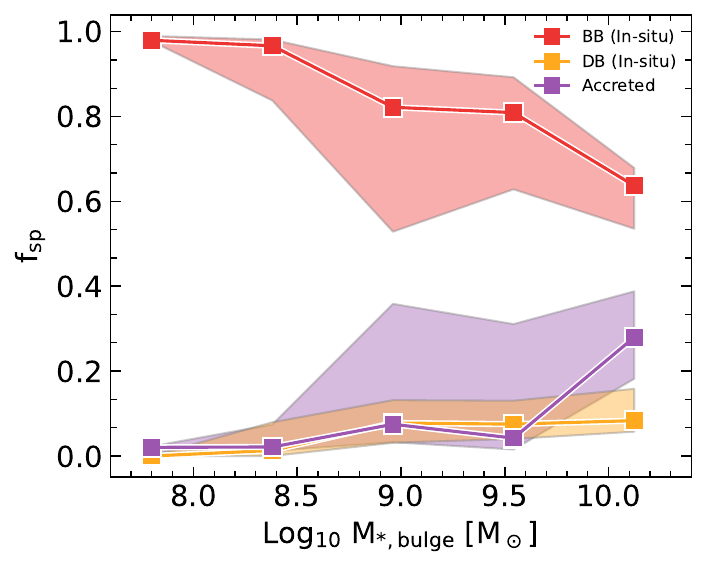}
         \caption{ Mass fraction of stellar populations contributing to the bulges. Stars formed in-situ (red dots and shaded regions), accreted stars (violet dots and shaded regions) and stars formed in the disk component and subsequently incorporated  into the bulge through secular evolution (yellow dots and shaded regions) are shown. The median values (dots and lines) and the 16$^{\rm th}$-84$^{\rm th}$ percentiles (shaded regions) of each of the different populations are depicted.}
         \label{fig:mass_bt_facc}
\end{figure}
\begin{figure}
   \centering        
   \includegraphics[width = 0.5\textwidth]{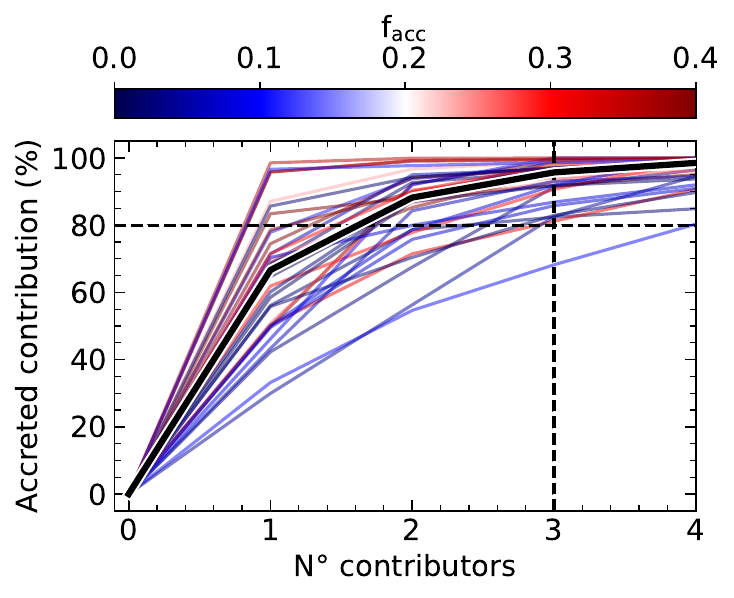}
         \caption{Cumulative contribution of satellites to the accreted stellar populations according to the number of major contributors. The cumulative tracks are color-coded by the accreted fraction of the corresponding bulges. As a reference, lines at 80\% and 3 contributors are depicted. The median trend of the contributions is included for reference (solid black line).} 
         \label{fig:contributors}
\end{figure}

\subsection{The simulated galaxies}\label{sec:decomposition}

We  selected  44 central galaxies defined as the most massive system within a virial halo. 

The stellar masses of simulated galaxies are in the range of  $[10^{7.6} - 10^{10.6}]$ $\Msun$. All selected galaxies are
resolved with more than 8000 stellar particles. We note that galaxies with stellar masses lower than 10$^9$ $\Msun$ are selected from the high resolution run only.

The dynamical components of the galaxies, bulge, disk and stellar halos were identified by using the AM-E method as explained in \citet{2012:Tissera}. This method defines  the circularity parameter, $\epsilon = J_z/J_{z,\rm max} (E_{\rm bind})$, where $J_z$ is the angular momentum  perpendicular to the rotational plane of each particle, and $J_{z,\rm max}$ is the maximum angular momentum possible for a given binding energy, $E_{\rm bind}$. Hence, stars on perfect circular orbits in a plane have $\epsilon$ = 1; conversely,
dispersion-dominated systems have $\epsilon \sim 0$. This, together with a threshold in $E_{\rm bind}$, allows the sorting of stellar particles into different components.  The bulge is defined by the stellar particles with E$_{\rm bind}$ lower than the minimum energy at 0.5 \ropt\footnote{The optical radius, \ropt, is defined as the one that enclosed 83 per cent of the stellar mass of  a galaxy.}, meaning that all stellar particles with lower binding energies are classified as part of the bulge. On the other hand, the disk component is defined as the stellar particles with $|\epsilon| > 0.5$ and  located within $2\times $\ropt. These criteria have been previously used to study \cielo~galaxies \citep{2025:Gonzalez,2025:Tissera,2025:Tapia}, so we kept them for consistency. With this classification, we can also estimate the bulge-to-total mass ratio, B/T. The selected galaxies have B/T within the range 0.20 to 0.85. Hence, our sample encompasses a broad variety of morphologies, with different relative importance of the bulge and disk components.

Regarding the formation channels of the stellar populations in the simulated bulge components, we followed all stellar particles back in time to their site of formation and classified them accordingly. We grouped  them as accreted if they were born from material bound to accreted satellites at their time of formation, and as in-situ if they were formed within the progenitor of a  given galaxy (i.e. within 1.5 \ropt).
The most massive system in a halo is defined as the progenitor at a given time. Furthermore, the in-situ stellar populations are also divided into two subgroups, depending on whether they formed within the bulge component, hereafter bulge-born (BB),  or the disk component,hereafter disk-born, (DB). In the latter case, these stars  migrated into the bulge, probably as a result of local instabilities \citep{2011:Bournaud}  or tidally-induced torques produced by nearby companions \citep{2013:Perez}.

Figure~\ref{fig:mass_bt_facc} shows, as a function of bulge mass, the stellar mass fractions, f$_{\rm sp}$, of three distinct populations in the bulge, separated by their formation channel, BB and DB  and accreted.  As can be seen, the predominant formation channel is the in-situ formation of stars from gas already present in the bulge (BB), with f$_{\rm sp}$ varying from nearly 1 down to 0.60. However, as we go to higher masses the contributions of accreted and DB increase slightly. In particular, $\sim$ 30\% of the galaxies in the sample tend to have accretion fractions (\facc) higher than \facc$=0.20$. Therefore, the two main formation channels we identify are bulge-born (BB) and accreted stars. However, some galaxies also show significant contributions from DB populations\footnote{We acknowledge that numerical resolution could prevent or weaken the formation of bars and that would have an impact on the fraction of stars affected by secular evolution in our intermediate resolution run \citep{2025:Fragkoudi} Hence, the fraction of DB could be considered as a lower limit. }, as illustrated in Fig.~\ref{fig:mass_bt_facc}. 

To better quantify the origin of the accreted stellar populations, we identified the satellites to which the accreted material was bound. This allows us to track the number of satellites that contribute to the formation of a given stellar population in the bulge. Figure~\ref{fig:contributors} shows the cumulative accreted contribution to the bulge as a function of the number of major contributors. To obtain the major satellite  contributors, we only include bulges with accreted masses higher than $10^6$ $\Msun$ and $10^7$$\Msun$ for the intermediate and high resolution simulations, respectively (i.e. if bulges have at least 50 accreted particles). This is done to reduce numerical noise due to low number particles. We note that our results  do not depend on this adopted numerical limit if it remains low, since we are mainly concerned about galaxies with significant accretion fraction.

As can be seen Fig.~\ref{fig:contributors}  while there is a variety of accreted fractions, most bulges reach at least 80\% of their accreted stars by the contribution of  two or three accreted satellites. There is a slight trend for bulges with \facc$<0.2$ to have a larger number of  contributors. This could be due to the accretion of stripped particles mainly at high redshift, where the assembly history is more chaotic. These small bulges do not have main accretion events.

\section{The stellar mass-metallicity relation of bulges}\label{sec:3}

\begin{figure}
   \centering        
   \includegraphics[width = 0.5\textwidth]{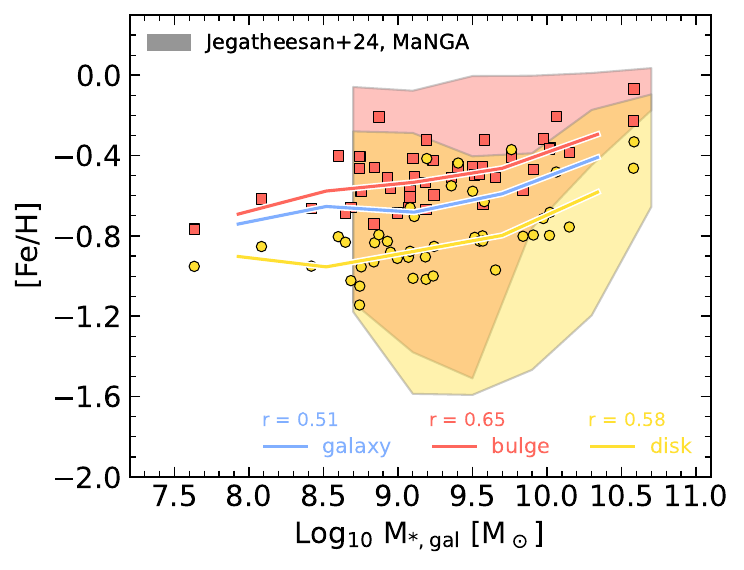}
         \caption{The \sMZR~of the bulges (red squares) and disks (yellow circles) of the \cielo~galaxies, defined by the median [Fe/H]. For comparison, the \cielo~galaxy \sMZR~is also included (light blue line). Additionally, observational trends reported by \citet{2024:Jegatheesan} for the bulges (red shaded regions) and disks (yellow shaded regions) are also included. Shaded regions are determined by the  $16^{\rm th}$-$84^{\rm th}$ percentiles.} 
         
         \label{fig:MZRcomponents}
\end{figure}

We estimated the median [Fe/H] of the stellar populations in the simulated bulges and disks and built the corresponding \sMZR\footnote{The estimated values are rescaled using the solar abundances from \citet{2019:Lodders}.}.
Figure~\ref{fig:MZRcomponents} displays these relations and that of the galaxies as a whole. The disk and the bulge components determine an \sMZR~with  similar shapes, but with different  level of enrichment. The Pearson correlation coefficients (r) show well-defined correlation for all cases (p-values are all lower than $p < 0.001$). As can be seen in Fig.~\ref{fig:MZRcomponents}, the bulges tend to be more enriched than the disks.

The simulated trends  are consistent with those from the MaNGA survey reported by \citet{2024:Jegatheesan} (shaded regions). We note that these authors applied BUDDI \citep{2017:Johnston} to  decompose  bulge and disk components using two-dimensional light profiles. This procedure is different from our bulge and disk decomposition (see Section 2.1). Hence, we do not aim at making a detail comparison but just to compare the global trends of the corresponding \sMZR. To match the enrichment levels between observations and simulations, we rescaled the observational data to match the median abundances of the \cielo~galaxies at M$_{*,\mathrm{gal}} \sim 10^{9.5} \Msun$. This corresponds to a downward shift in [Fe/H] of 0.21 dex. We make this correlations since the methods to determine the abundances in observations and simulations are different.

As shown in Fig.~\ref{fig:MZRcomponents}, bulges tend to be more enriched than the disk components, although in some cases both exhibit similar levels of enrichment. This trend is expected considering the bulges host stellar populations born in the central regions which are very dense, and hence experienced strong starbursts. The disks continued to accrete gas and form stars in a more steady fashion. Nevertheless, there is some overlap with a subsample of galaxies that also exhibit enriched disks in agreement with observations.

\begin{figure*}
   \centering          \includegraphics[width = 0.95\textwidth]{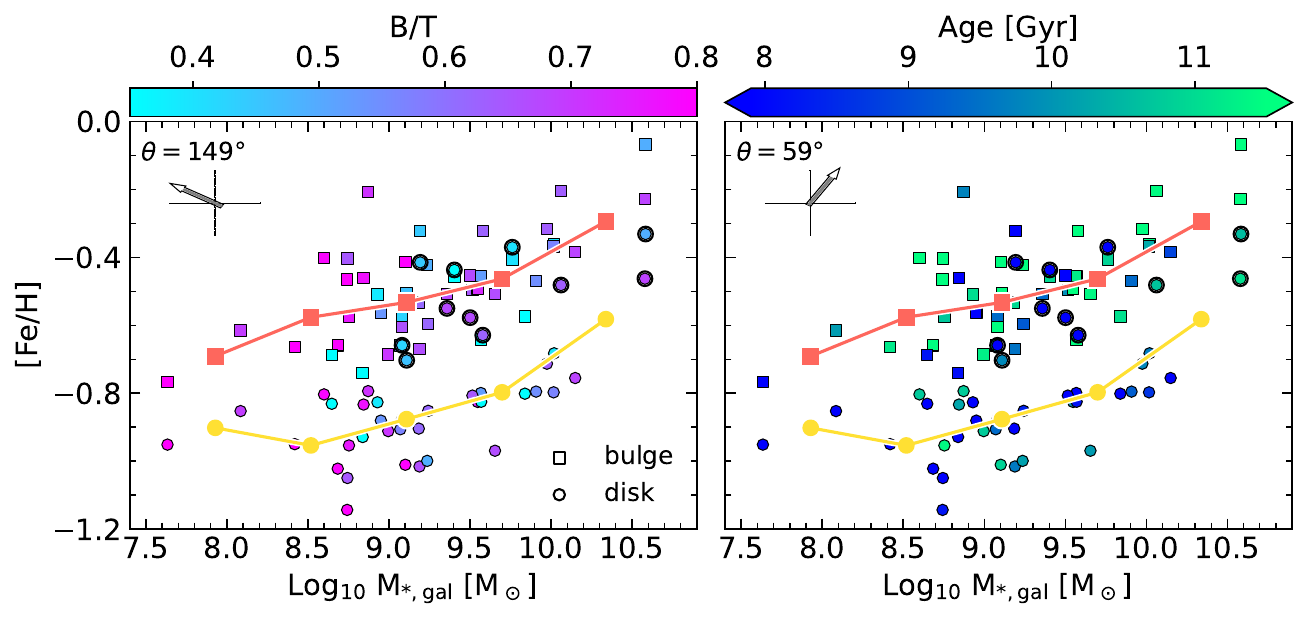}
         \caption{[Fe/H] for the bulge (squares) and disk (circles) components as a function of the galaxy stellar mass. The solid lines denote the median trends for both components(orange and yellow, respectively). The plane is color-coded by bulge-total-ratio (left panel) and median age of the corresponding components (right panel). The disk components  with an excess of [Fe/H], respect to the disk \sMZR, are highlighted. To quantify third dependences of the bulge, Partial Correlation Coefficients analysis \citep[PCC,][]{2020:Bluck} angles are included.}\label{fig:MZR_btage}
\end{figure*}

To compare the galaxy morphology and ages of the components, Fig.~\ref{fig:MZR_btage} shows the \sMZR~of the bulge and disk color-coded by the galaxy B/T in the left panel, and the median age of the corresponding component in the right one. From the left panel we see a weak secondary correlations with B/T,  so that   lower mass galaxies  tend to be more bulge-dominated. To quantify this dependence, we estimated the correlation angle using Partial Correlation Coefficients analysis \citep[PCC,][]{2020:Bluck}. For the third correlation with B/T, the angle is $149$  with a bootstrap error of $20$, suggesting a decrease of B/T for an increasing bulge \mstar. 
The right panel of Fig.~\ref{fig:MZR_btage} shows that  more massive bulges tend  to be slightly older. The PCC analysis supports this with an angle  of $59 \pm 44$.  Considering the large bootstrap error, this angle should be taken just as a reference of the relative direction of the third correlation.

\begin{figure*}
   \centering        \includegraphics[width = 0.95\textwidth]{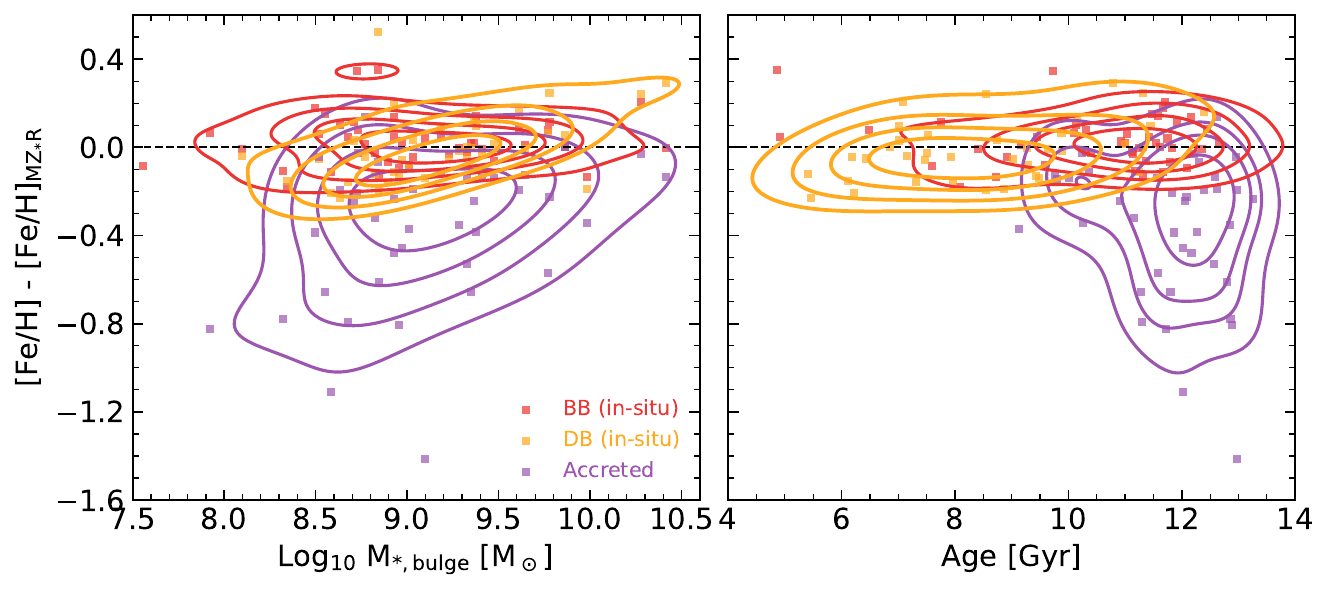}
         \caption{Offset between the metallicity of a given stellar population and the metallicity predicted by the bulge \sMZR~for the corresponding bulge mass as $\rm  [Fe/H] -[Fe/H]_{\rm MZ_*R}$. We considered  accreted (purple), disk born (gold) and bulge born (orange) stellar populations. Left panel: metallicity offset as a function of the stellar mass of the bulge (left panel) and the median age of the stellar population itself (right panel). The lines correspond to the density countours at 20\%, 40\%, 60\% and 80\% of each distribution. }\label{fig:MZR_origin}
\end{figure*}

Regarding the disk components, the median \sMZR~is determined by the bulk of the systems, and the metallicity is about 0.34 dex lower than that of the bulges. In the case of MaNGA, the disk \sMZR~is about 0.6 dex more metal-poor than the bulges. However, some simulated  disks exhibit higher metallicities. These systems lie along the bulge \sMZR\ (black contours). Some of them  display  prominent disk components populated by  younger stellar populations. They have  median  B/T of $ 0.49^{0.67}_{0.42}$ and median ages of $7.00^{10.60}_{5.31}$ Gyr. Meanwhile the bulk of the disk components have median B/T of $0.65^{0.79}_{0.40}$ and median ages of $8.40^{10.23}_{5.21}$ Gyr (the lower and upper numbers correspond to the $16^{\rm th}$ and $84^{\rm th}$ percentiles). Those galaxies with younger disks continue their star formation probably due to continuous gas accretion, which also contributes to the formation of the disks \citep{Yu:2023}. However, our analysis shows an important variation of properties that  reflect the complex formation processes of bulges. In the preceding analysis, disk properties have been  included for completeness, but the analysis hereafter focuses exclusively on the bulge components.

 \begin{figure*}
   \centering        \includegraphics[width = 0.95\textwidth]{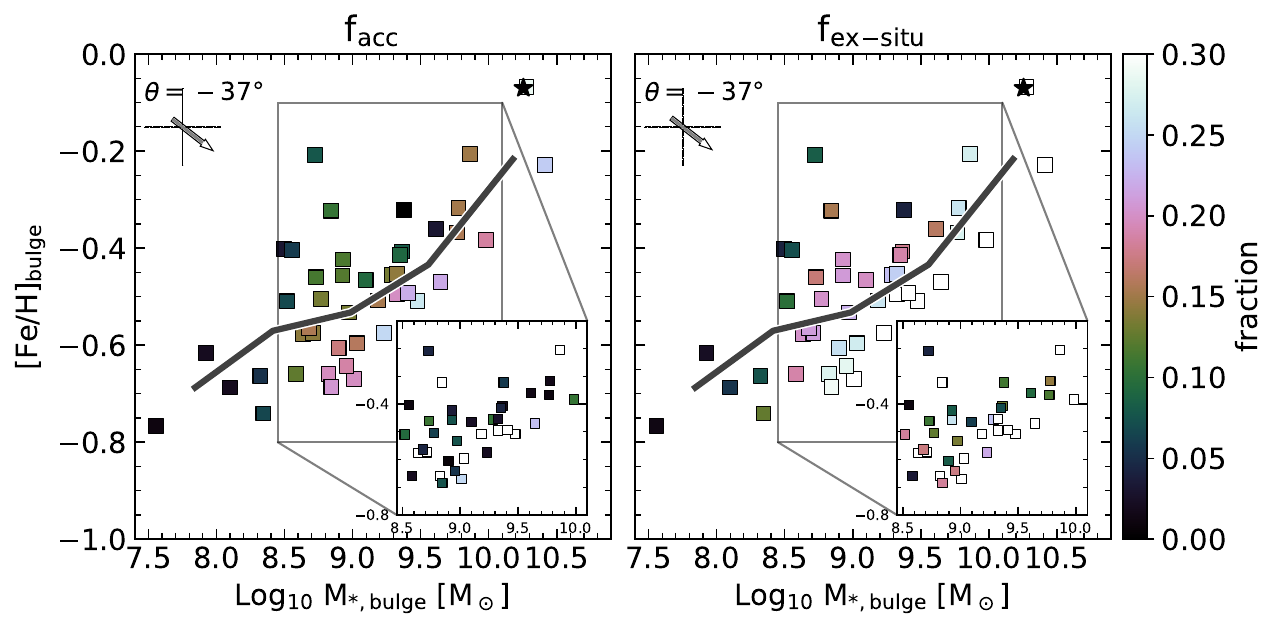} \caption{Mass-metallicity relation for the simulated bulges color-coded by accreted fraction , $f_{\rm acc}$ (left panel) and the accreted plus the DB fraction, \fout~(right panel). The LOESS2D smoothing algorithm was applied to highlight the trend \citep{2013:Cappellari}. The inset figure shows the distribution without LOESS2D for the mass range of interest  for comparison (i.e. the actual \facc, \fout~values.). As a reference, the Milky Way bulge measurements of stellar mass \citep{2017:Portail} and median [Fe/H] \citep{2017:Zocalli} are included (black star)}.
   \label{fig:mzr_facc}
\end{figure*}

To further analysis the properties of the three stellar population that form the bulges, we estimated the offset between the metallicity of a given stellar population and the metallicity predicted by the bulge \sMZR~for the corresponding bulge mass as $\rm  [Fe/H] -[Fe/H]_{\rm MZ_*R}$. Figure~\ref{fig:MZR_origin} compares the level of enrichment and ages of BB (red), DB (yellow) and accreted (purple) stars. The left panel of Fig.~\ref{fig:MZR_origin} shows the offset of metallicity as a function of the bulge stellar mass, while the right figure displays the offsets as a function of the median age of each stellar population.  An important aspect of this estimation is that we only consider a stellar population if it is resolved with more than 50 stellar particles, in order to minimize numerical noise.

We find that the metallicity of the in-situ stellar populations is closer to the bulges \sMZR~as expected, since they are the predominant population. The DB populations exhibit enrichment levels comparable to the BB ones, although they display a mild trend of increasing metallicity with stellar mass. This is likely due to the more enriched ISM of higher-mass galaxies, where these stars originated before migrating to the bulge. Finally, the accreted stellar populations are in general less enriched, compared to the in-situ populations. This is expected since the major contributors of these accreted stellar populations to the bulges are low mass satellite galaxies, with less enriched ISM. In Fig.~\ref{fig:mzr_contributors} we show the MZR of the three main satellite contributors to the bulges together with the bulge MZR (see Sec.~\ref{sec:ap3}). From this figure, it is also clear that less massive satellites entered the virial radius at higher lookback times and hence, they did not have time for chemical evolution to take place. 

The right panel of Fig.~\ref{fig:MZR_origin} shows that the accreted populations tend to be as old as the in-situ bulge-born stars. The difference between these two populations is the level of enrichment. Conversely, the DB stars present a wider range of ages, but in general they are younger than the other populations. This is expected, as disks typically form at later evolutionary stages and continue forming stars over extended periods. 
\begin{figure*}
   \centering        \includegraphics[width = 0.95\textwidth]{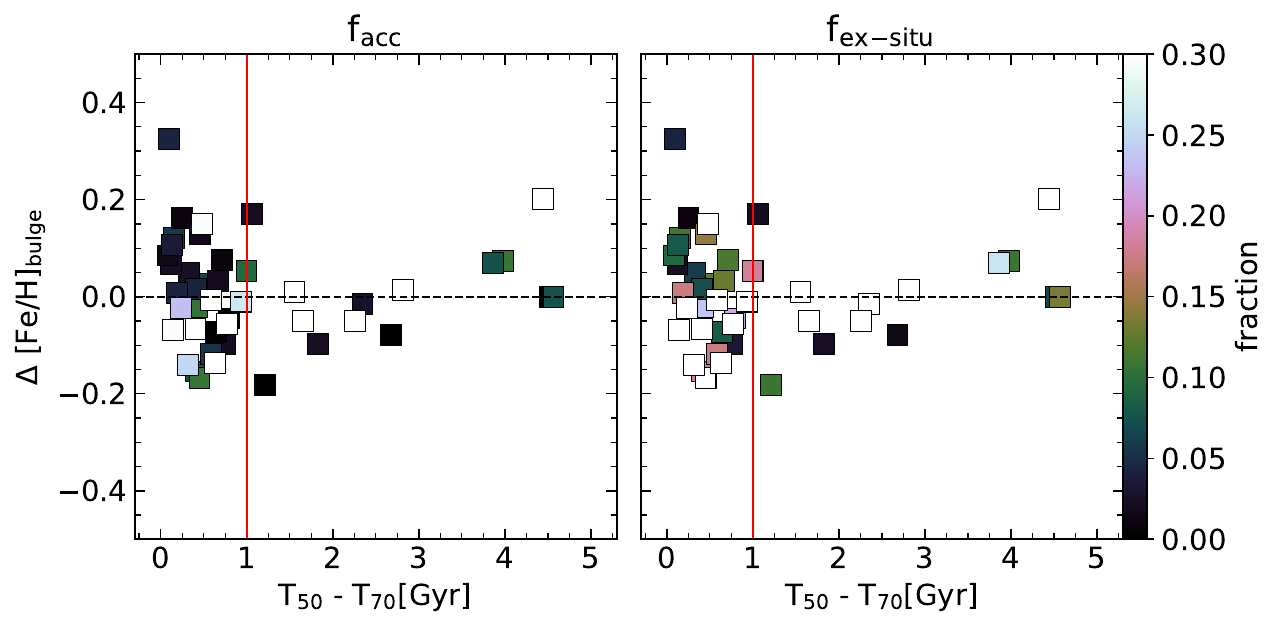}\caption{Offset of [Fe/H] with respect to the median \sMZR~, $\Delta \rm [Fe/H]_{\rm bulges}$, as a function of the difference between the two formation times, T$_{50}$ and T$_{70}$. We show the secondary dependence of the offset on the \facc~(right panel), and \fout, defined as the sum of \facc~and the fraction of DB stars (left panel). The separation in fast and bursty formation history is given by a threshold in T$_{50}$ - T$_{70}  \sim 1$  (red vertical line).}
   \label{offsets}
\end{figure*}

According to these results,   accreted stars  decrease the overall metallicity of the bulge, depending on the relevance of their contribution. To quantify this effect, the left panel of Fig.~\ref{fig:mzr_facc} presents the bulge \sMZR~color-coded by the \facc. We applied  the LOESS2D algorithm to highlight the dependence on \facc~\citep{2013:Cappellari}. For bulges with \mstar~lower than $10^{8.5} \Msun$, there is no clear correlation between the metallicity relation and the accretion fractions. However, the number of galaxies in this mass range  is too low to draw a conclusion. Conversely, for \mstar$ > 10^{8.5} \Msun$ we do find a dependence on \facc, so that at a given bulge mass, those with lower metallicity tend to have larger accreted contribution.

As previously noted, some bulges also exhibit contributions from DB stellar populations that migrate toward the central region. Hence, we also define \fout~as the combined fraction of accreted and DB stars. The right panel of Fig.~\ref{fig:mzr_facc} shows the \sMZR~as a function of  \fout. 
Adding the DB contribution reinforces the trend initially seen with \facc~alone, indicating that bulges with lower metallicities are associated with larger fractions of stars formed outside the bulge.

The PCC angle shows the direction at which the third correlation, in this case with \facc~and \fout, increases with the bulge \sMZR. The correlation angles, $\theta$, present similar values for both relations, with bootstrap errors lower than 4°. 
We stress the fact that this trend is not a direct correlation between bulge mass and total galaxy mass. Although bulge mass does correlate with stellar mass along the \sMZR~(i.e. more massive bulges are typically found in more massive galaxies), this relation does not account for the dispersion at a given bulge mass, as shown in Fig.~\ref{fig:mzr_galmass}.

The trend to have lower metallicities in bulges with higher \facc, at a given stellar mass,  is clear.  More massive galaxies are expected to have more  mergers, increasing the probability to have a higher contribution of accreted stars in the bulge \citep{2024:Angeloudi}. Meanwhile, the contribution of DB stars produce a similar effect because these stars are less enriched, in general,  as  galaxies show negative metallicity profiles \citep{2025:Tapia}.

We also include in both panels of Fig.~\ref{fig:mzr_facc},  a zoom-in view of the intermediate mass range ($10^{8.5}-10^{10} \Msun$) without LOESS2D smoothing, to assess that there are not spurious results originated by using this smoothing procedure. First we can see that results hold, indicating that the less metal-rich galaxies tend to have higher accretion fractions. Second, there are outliers,  exhibiting higher \facc~and higher metallicity (i.e. when a very massive merger occurred)  for example. Together, these findings highlight the diversity of bulge assembly pathways.

In summary, we detect a \sMZR~for bulges, which follows the overall increasing trend of the galaxy \sMZR~but at higher enrichment levels. We find that part of the scatter in this relation is linked to the assembly history of bulges, such that those with larger contributions from accreted and DB populations tend to shift toward lower metallicity at a given bulge mass. Nevertheless, some systems deviate from this trend, and in the next section we discuss the origin of this diversity.

\section{Assembly histories of bulges and the \sMZR}\label{sec:assembly}

In order to further analyze the formation histories of the bulges,  we estimated the cumulated stellar mass as a function of age, and calculated the lookback time at which 50\% and 70\% of the stellar population in the bulges was formed,  T$_{50}$ and  T$_{70}$ respectively.  In Fig.~\ref{offsets} we show the metallicity offsets of individual bulges  with respect to the \sMZR~, $\Delta \rm [Fe/H]_{\rm bulges}$, as a function of $\rm T_{50} - T_{70}$. The left panel shows the third dependence on  \facc~while the right panel shows the dependence on \fout.

The difference between these times provides a way to discriminate between bulges with fast formation histories. From this figure we can see that most of the bulges are concentrated at   $\rm T_{50} - T_{70} \leq  1 $ Gyr, implying  $\rm T_{50} \sim T_{70}$. We adopt this age threshold to define bulge with fast formation. The rest of the system with with bursty or extended formation are assumed to have  $\rm T_{50} - T_{70} >  1$ Gyr. For our sample, we estimate that $\sim$ 67\% of bulges are consistent with  a fast formation path.

  As can be seen in the left panel of Fig.~\ref{offsets}, bulges with fast formation and high accretion tend to have negative offsets. However, there are also bulges that fall below the median \sMZR~that  have  fast formation and low accretion.  
  When we include both fractions together, as shown the right panel of Fig.~\ref{offsets},  we can see that the trend is stronger and most fast forming bulges with negative offsets have high \fout. This can be understood because bulges with low accretion, \facc$< 0.2$, and fast formation have a median \fmig~of $11.9^{19.7}_{7.2}$\% for negative offsets and $5.8^{10.9}_{0.2}$\% for positive offsets (lower and upper numbers represent the $16^{\rm th}$ and $84^{\rm th}$ percentiles). The different \fmig~of  offsets from the bulge \sMZR~suggest that migration of stellar populations from the disk could also affect the median [Fe/H].
  Some of them received old and low metallicity stars while other have latter contributions of younger, more metal-rich stars.  There is only an exception, the most massive bulge with a fast formation and a positive offset, which had a major merger that contributed with high metallicity stars.
  
  On the other hand,  bulges with bursty star formation and an excess of [Fe/H] tend to have a larger fraction of \fout, which is reinforced by the DB stars. We find that the \fmig~for low accretion is $5.8^{13.4}_{0.0}$\% for negative offsets and $0.0^{12.6}_{0.0}$\% for positive offsets. For high accretion \fmig~is $5.7^{6.5}_{4.9}$\% for negative offsets and $8.1^{8.4}_{6.1}$\% for positive offsets. In this case, the larger contribution of migrated stars could have brought more enriched material since these bulges have an extended formation history. In fact, the DB stars are in median $\sim 2$ Gyr younger for the bursty scenarios, when compared to the fast forming ones, suggesting that they are being form in a more enriched ISM. However, caution should be taken regarding the results on  bursty bulges because of the low number of systems. 
  
In summary, we find that bulges regarding of their formation history determine a \sMZR~and that the dispersion of this \sMZR~ is affected by both, stellar accretion from satellites and migration from the disk, so that at a given bulge mass, those with larger \fout~tend to be  less metal-rich.

\section{Conclusions}
We performed a chemo-dynamical analysis of the bulges of 44 central galaxies of the \cielo~simulations. We built up the \sMZR~of bulges and showed that the dispersion at a given bulge mass could be related to the accretion history.

Our main results can be summarized as follows,
\begin{itemize}
    \item We found three channel of formation of the stellar populations in the \cielo~bulges. Most of the stellar populations are formed in-situ. However, about 33\% of the simulated bulges  have \facc$ > 0.2$. The third channel of formation are stars born in the disk component and later, acquired by the bulges via secular evolution. They can contribute up to about 32\% of the stellar populations in the more massive bulges. A trend with the bulge mass is found so that bulges with greater mass have a more significant contribution of the last two channels.

    \item We found that bulges follow a \sMZR~similar in shape to the galaxy \sMZR, but, at least, for the \cielo~simulations, with a higher level of enrichment of about $\sim$ 0.13 dex. The simulated bulge \sMZR~is in global agreement with results from the MaNGA survey and is driven mainly by the in-situ (BB) stellar component.

    \item Accreted stars are found to be older and to have lower metallicity than in-situ stars, as expected, since they formed in galaxies with lower mass than the central ones. 
    Our results show that most of  the accreted material is provided by between two or three satellites utmost. The main satellite contributors determine a MZR which is steeper and show a lower level of enrichment than the bulge \sMZR. 
    
    \item DB stellar populations tend to be younger than BB stars and slightly more enriched in more massive systems, consistent with having been  formed from a more chemical enriched interstellar medium in the corresponding disk components. 
    
    \item Bulges with higher \facc~tend to have lower metallicity at a given bulge mass. This trend is reinforced by adding \fmig. Hence, the bulge \sMZR~stores relevant information of the assembly history of the bulges, which is reflected in its dispersion  of the \sMZR~at a given stellar mass. This trend resembled the behavior reported for the \sMZR~of stellar halos \citep[][Gonzalez-Jara in prep]{DSouza_2018}, where the dispersion is dominated by the most massive contributing satellite. In the case of the bulges, most of the stars are formed in situ, but the dispersion at a given bulge mass is  modulated by the ex-situ stellar contribution.

\end{itemize}

These results reflects that the formation of the bulges  involves different formation paths that yields different chemical enrichment. However, a \sMZR~is clear at place driven mainly by the in-situ stellar population, which is the dominating component,  but with  a dispersion modulated by the impact of other channels of formation related with stars formed ex-situ.

\begin{acknowledgements}
    We thank the anonymous referee for the constructive report, which helped to improve this paper. We thank E. Johnston and K. Jegatheesan to share their database from the MaNGA survey. We acknowledge ANID Basal Project FB210003. IM acknowledges funding by ANID
(Beca Magíster Nacional, Folio 22241737). 
      PBT acknowledges partial funding by Fondecyt-ANID 1240465/2024. This project has received funding from the European Union Horizon 2020 Research and Innovation Programme under the Marie Sklodowska-Curie grant agreement No 734374- LACEGAL. JGJ acknowledges funding by ANID (Beca Doctorado Nacional, Folio
21210846). VPM acknowledges funding by ANID (Beca Magíster Nacional, Folio 22241063). This project used the Ladgerda Cluster (Fondecyt 1200703/2020 hosted at the Institute for Astrophysics, Chile), the NLHPC (Centro de Modelamiento Matem\'atico, Chile), Geryon clusters (Center for Astrophysics, CATA, Chile), and the Barcelona Supercomputer Center (Spain). 
\end{acknowledgements}


\bibliographystyle{aa}
\bibliography{references.bib}

@ARTICLE{rojas2019,
       author = {{Rojas-Arriagada}, A. and {Zoccali}, M. and {Schultheis}, M. and {Recio-Blanco}, A. and {Zasowski}, G. and {Minniti}, D. and {J{\"o}nsson}, H. and {Cohen}, R.~E.},
        title = "{The bimodal [Mg/Fe] versus [Fe/H] bulge sequence as revealed by APOGEE DR14}",
      journal = {\aap},
         year = 2019,
        month = jun,
       volume = {626},
          eid = {A16},
        pages = {A16},
          doi = {10.1051/0004-6361/201834126},
archivePrefix = {arXiv},
       eprint = {1905.01364},
 primaryClass = {astro-ph.GA},
       adsurl = {https://ui.adsabs.harvard.edu/abs/2019A&A...626A..16R}
}

@ARTICLE{lequeux1979,
       author = {{Lequeux}, J. and {Peimbert}, M. and {Rayo}, J.~F. and {Serrano}, A. and {Torres-Peimbert}, S.},
        title = "{Chemical Composition and Evolution of Irregular and Blue Compact Galaxies}",
      journal = {\aap},
         year = 1979,
        month = dec,
       volume = {80},
        pages = {155},
       adsurl = {https://ui.adsabs.harvard.edu/abs/1979A&A....80..155L}
}

@ARTICLE{skillman1989,
       author       = {Skillman, E. D. and Kennicutt, R. C. and Hodge, P. W.},
       title        = {Oxygen abundances in nearby dwarf irregular galaxies},
       journal      = {ApJ},
       year         = {1989},
       volume       = {347},
       pages        = {875--902},
       doi          = {10.1086/168180},
       adsurl       = {https://ui.adsabs.harvard.edu/abs/1989ApJ...347..875S}
}

@ARTICLE{zaritsky1994,
       author       = {Zaritsky, D. and Kennicutt, R. C. and Huchra, J. P.},
       title        = {H II regions and the abundance properties of spiral galaxies},
       journal      = {ApJ},
       year         = {1994},
       volume       = {420},
       pages        = {87--109},
       doi          = {10.1086/173544},
       adsurl       = {https://ui.adsabs.harvard.edu/abs/1994ApJ...420...87Z}
}

@ARTICLE{DSouza_2018,
       author = {{D'Souza}, Richard and {Bell}, Eric F.},
        title = "{The masses and metallicities of stellar haloes reflect galactic merger histories}",
      journal = {\mnras},
         year = 2018,
        month = mar,
       volume = {474},
       number = {4},
        pages = {5300-5318},
          doi = {10.1093/mnras/stx3081},
archivePrefix = {arXiv},
       eprint = {1705.08442},
 primaryClass = {astro-ph.GA},
       adsurl = {https://ui.adsabs.harvard.edu/abs/2018MNRAS.474.5300D}
}

@ARTICLE{campas2021,
       author       = {Camps-Fariña, A. and Sánchez, S. F. and Lacerda, E. A. D. and López-Cobá, C. and López-Sánchez, Á. R. and Galbany, L. and Rosales-Ortega, F. F. and Barrera-Ballesteros, J. K. and Espinosa-Ponce, C. and Sánchez-Menguiano, L. and Cano-Díaz, M. and Ibarra-Medel, H. J. and Zibetti, S. and García-Benito, R. and Cid Fernandes, R. and González Delgado, R. M. and Pérez, E. and Walcher, C. J.},
       title        = {Evolution of the chemical enrichment and the mass–metallicity relation in CALIFA galaxies},
       journal      = {Monthly Notices of the Royal Astronomical Society},
       year         = {2021},
       month        = jun,
       volume       = {504},
       number       = {3},
       pages        = {3478--3499},
       doi          = {10.1093/mnras/stab1019},
       adsurl       = {https://ui.adsabs.harvard.edu/abs/2021MNRAS.504.3478C}
}

@ARTICLE{tissera2006,
       author = {{Tissera}, P.~B. and {Smith Castelli}, A.~V. and {Scannapieco}, C.},
        title = "{Interactions, mergers and the fundamental mass relations of galaxies}",
      journal = {\aap},
         year = 2006,
        month = aug,
       volume = {455},
       number = {1},
        pages = {135-143},
          doi = {10.1051/0004-6361:20054747},
archivePrefix = {arXiv},
       eprint = {astro-ph/0603707},
 primaryClass = {astro-ph},
       adsurl = {https://ui.adsabs.harvard.edu/abs/2006A&A...455..135T}
}

@ARTICLE{gonzalezdelgado2014,
       author       = {González Delgado, R. M. and Cid Fernandes, R. and García-Benito, R. and Pérez, E. and de Amorim, A. L. and Sánchez, S. F. and Husemann, B. and Iglesias-Páramo, J. and Márquez, I. and Mollá, M. and Mast, D. and Kehrig, C. and Papaderos, P. and Vale Asari, N. and Walcher, J. and Alves, J. and Bland-Hawthorn, J. and Galbany, L. and Rosales-Ortega, F. and van de Ven, G. and Vilchez, J. M. and Wisotzki, L. and Ziegler, B.},
       title        = {Insights on the Stellar Mass–Metallicity Relation from the CALIFA Survey},
       journal      = {ApJ Letters},
       year         = {2014},
       month        = jul,
       volume       = {791},
       number       = {1},
       pages        = {L16},
       doi          = {10.1088/2041-8205/791/1/L16},
       adsurl       = {https://ui.adsabs.harvard.edu/abs/2014ApJ...791L..16G}
}

@ARTICLE{romano2023,
       author = {{Romano}, Donatella and {Ferraro}, Francesco R. and {Origlia}, Livia and {Portegies Zwart}, Simon and {Lanzoni}, Barbara and {Crociati}, Chiara and {Massari}, Davide and {Dalessandro}, Emanuele and {Mucciarelli}, Alessio and {Rich}, R. Michael and {Calura}, Francesco and {Matteucci}, Francesca},
        title = "{Modeling the Chemical Enrichment History of the Bulge Fossil Fragment Terzan 5}",
      journal = {\apj},
         year = 2023,
        month = jul,
       volume = {951},
       number = {2},
          eid = {85},
        pages = {85},
          doi = {10.3847/1538-4357/acd8ba},
archivePrefix = {arXiv},
       eprint = {2305.15355},
 primaryClass = {astro-ph.GA},
       adsurl = {https://ui.adsabs.harvard.edu/abs/2023ApJ...951...85R}
}

@ARTICLE{2024:Jegatheesan,
       author = {{Jegatheesan}, Keerthana and {Johnston}, Evelyn J. and {H{\"a}u{\ss}ler}, Boris and {Nedkova}, Kalina V.},
        title = "{BUDDI-MaNGA. III. The mass-assembly histories of bulges and discs of spiral galaxies}",
      journal = {\aap},
         year = 2024,
        month = apr,
       volume = {684},
          eid = {A32},
        pages = {A32},
          doi = {10.1051/0004-6361/202347372},
archivePrefix = {arXiv},
       eprint = {2402.00959},
 primaryClass = {astro-ph.GA},
       adsurl = {https://ui.adsabs.harvard.edu/abs/2024A&A...684A..32J}
}

@ARTICLE{2017:Zocalli,
       author = {{Zoccali}, M. and {Vasquez}, S. and {Gonzalez}, O.~A. and {Valenti}, E. and {Rojas-Arriagada}, A. and {Minniti}, J. and {Rejkuba}, M. and {Minniti}, D. and {McWilliam}, A. and {Babusiaux}, C. and {Hill}, V. and {Renzini}, A.},
        title = "{The GIRAFFE Inner Bulge Survey (GIBS). III. Metallicity distributions and kinematics of 26 Galactic bulge fields}",
      journal = {\aap},
         year = 2017,
        month = mar,
       volume = {599},
          eid = {A12},
        pages = {A12},
          doi = {10.1051/0004-6361/201629805},
archivePrefix = {arXiv},
       eprint = {1610.09174},
 primaryClass = {astro-ph.GA},
       adsurl = {https://ui.adsabs.harvard.edu/abs/2017A&A...599A..12Z}
}

@ARTICLE{2020:Breda,
       author = {{Breda}, Iris and {Papaderos}, Polychronis and {Gomes}, Jean-Michel},
        title = "{Indications of the invalidity of the exponentiality of the disk within bulges of spiral galaxies}",
      journal = {\aap},
         year = 2020,
        month = aug,
       volume = {640},
          eid = {A20},
        pages = {A20},
          doi = {10.1051/0004-6361/202037889},
archivePrefix = {arXiv},
       eprint = {2006.02307},
 primaryClass = {astro-ph.GA},
       adsurl = {https://ui.adsabs.harvard.edu/abs/2020A&A...640A..20B}
}

@ARTICLE{2019:Gargiulo,
       author = {{Gargiulo}, Ignacio D. and {Monachesi}, Antonela and {G{\'o}mez}, Facundo A. and {Grand}, Robert J.~J. and {Marinacci}, Federico and {Pakmor}, R{\"u}diger and {White}, Simon D.~M. and {Bell}, Eric F. and {Fragkoudi}, Francesca and {Tissera}, Patricia},
        title = "{The prevalence of pseudo-bulges in the Auriga simulations}",
      journal = {\mnras},
         year = 2019,
        month = nov,
       volume = {489},
       number = {4},
        pages = {5742-5763},
          doi = {10.1093/mnras/stz2536},
archivePrefix = {arXiv},
       eprint = {1907.02082},
 primaryClass = {astro-ph.GA},
       adsurl = {https://ui.adsabs.harvard.edu/abs/2019MNRAS.489.5742G}
}

@ARTICLE{2005:Springel,
       author = {{Springel}, Volker},
        title = "{The cosmological simulation code GADGET-2}",
      journal = {\mnras},
         year = 2005,
        month = dec,
       volume = {364},
       number = {4},
        pages = {1105-1134},
          doi = {10.1111/j.1365-2966.2005.09655.x},
archivePrefix = {arXiv},
       eprint = {astro-ph/0505010},
 primaryClass = {astro-ph},
       adsurl = {https://ui.adsabs.harvard.edu/abs/2005MNRAS.364.1105S}
}

@ARTICLE{2003:Springel,
       author = {{Springel}, Volker and {Hernquist}, Lars},
        title = "{Cosmological smoothed particle hydrodynamics simulations: a hybrid multiphase model for star formation}",
      journal = {\mnras},
         year = 2003,
        month = feb,
       volume = {339},
       number = {2},
        pages = {289-311},
          doi = {10.1046/j.1365-8711.2003.06206.x},
archivePrefix = {arXiv},
       eprint = {astro-ph/0206393},
 primaryClass = {astro-ph},
       adsurl = {https://ui.adsabs.harvard.edu/abs/2003MNRAS.339..289S}
}

@ARTICLE{2022:Rodriguez,
       author = {{Rodr{\'\i}guez}, S. and {Garcia Lambas}, D. and {Padilla}, N.~D. and {Tissera}, P. and {Bignone}, L. and {Dominguez-Tenreiro}, R. and {Gonzalez}, R. and {Pedrosa}, S.},
        title = "{Satellite galaxies in groups in the CIELO Project I. Gas removal from galaxies and its re-distribution in the intragroup medium}",
      journal = {\mnras},
         year = 2022,
        month = aug,
       volume = {514},
       number = {4},
        pages = {6157-6172},
          doi = {10.1093/mnras/stac1377},
archivePrefix = {arXiv},
       eprint = {2205.06886},
 primaryClass = {astro-ph.GA},
       adsurl = {https://ui.adsabs.harvard.edu/abs/2022MNRAS.514.6157R}
}

@ARTICLE{2005:Scannapieco,
       author = {{Scannapieco}, C. and {Tissera}, P.~B. and {White}, S.~D.~M. and {Springel}, V.},
        title = "{Feedback and metal enrichment in cosmological smoothed particle hydrodynamics simulations - I. A model for chemical enrichment}",
      journal = {\mnras},
         year = 2005,
        month = dec,
       volume = {364},
       number = {2},
        pages = {552-564},
          doi = {10.1111/j.1365-2966.2005.09574.x},
archivePrefix = {arXiv},
       eprint = {astro-ph/0505440},
 primaryClass = {astro-ph},
       adsurl = {https://ui.adsabs.harvard.edu/abs/2005MNRAS.364..552S}
}

@ARTICLE{2006:Scannapieco,
       author = {{Scannapieco}, C. and {Tissera}, P.~B. and {White}, S.~D.~M. and {Springel}, V.},
        title = "{Feedback and metal enrichment in cosmological SPH simulations - II. A multiphase model with supernova energy feedback}",
      journal = {\mnras},
         year = 2006,
        month = sep,
       volume = {371},
       number = {3},
        pages = {1125-1139},
          doi = {10.1111/j.1365-2966.2006.10785.x},
archivePrefix = {arXiv},
       eprint = {astro-ph/0604524},
 primaryClass = {astro-ph},
       adsurl = {https://ui.adsabs.harvard.edu/abs/2006MNRAS.371.1125S}
}

@article{2012:Tissera,
    author = {Tissera, Patricia B. and White, Simon D. M. and Scannapieco, Cecilia},
    title = "{Chemical signatures of formation processes in the stellar populations of simulated galaxies}",
    journal = {Monthly Notices of the Royal Astronomical Society},
    volume = {420},
    number = {1},
    pages = {255-270},
    year = {2012},
    month = {01},
    issn = {0035-8711},
    doi = {10.1111/j.1365-2966.2011.20028.x},
    url = {https://doi.org/10.1111/j.1365-2966.2011.20028.x},
    eprint = {https://academic.oup.com/mnras/article-pdf/420/1/255/18456075/mnras0420-0255.pdf},
}

@ARTICLE{2010:Fisher,
       author = {{Fisher}, David B. and {Drory}, Niv},
        title = "{Bulges of Nearby Galaxies with Spitzer: Scaling Relations in Pseudobulges and Classical Bulges}",
      journal = {\apj},
         year = 2010,
        month = jun,
       volume = {716},
       number = {2},
        pages = {942-969},
          doi = {10.1088/0004-637X/716/2/942},
archivePrefix = {arXiv},
       eprint = {1004.5393},
 primaryClass = {astro-ph.CO},
       adsurl = {https://ui.adsabs.harvard.edu/abs/2010ApJ...716..942F}
}

@ARTICLE{2004:Kormendy,
       author = {{Kormendy}, John and {Kennicutt}, Robert C., Jr.},
        title = "{Secular Evolution and the Formation of Pseudobulges in Disk Galaxies}",
      journal = {\araa},
         year = 2004,
        month = sep,
       volume = {42},
       number = {1},
        pages = {603-683},
          doi = {10.1146/annurev.astro.42.053102.134024},
archivePrefix = {arXiv},
       eprint = {astro-ph/0407343},
 primaryClass = {astro-ph},
       adsurl = {https://ui.adsabs.harvard.edu/abs/2004ARA&A..42..603K}
}

@INPROCEEDINGS{2016:Kormendy,
       author = {{Kormendy}, John},
        title = "{Elliptical Galaxies and Bulges of Disc Galaxies: Summary of Progress and Outstanding Issues}",
    booktitle = {Galactic Bulges},
         year = 2016,
       editor = {{Laurikainen}, Eija and {Peletier}, Reynier and {Gadotti}, Dimitri},
       series = {Astrophysics and Space Science Library},
       volume = {418},
        month = jan,
        pages = {431},
          doi = {10.1007/978-3-319-19378-6_16},
archivePrefix = {arXiv},
       eprint = {1504.03330},
 primaryClass = {astro-ph.GA},
       adsurl = {https://ui.adsabs.harvard.edu/abs/2016ASSL..418..431K}
}

@ARTICLE{2023:Debattista,
       author = {{Debattista}, Victor P. and {Liddicott}, David J. and {Gonzalez}, Oscar A. and {Beraldo e Silva}, Leandro and {Amarante}, Jo{\~a}o A.~S. and {Lazar}, Ilin and {Zoccali}, Manuela and {Valenti}, Elena and {Fisher}, Deanne B. and {Khachaturyants}, Tigran and {Nidever}, David L. and {Quinn}, Thomas R. and {Du}, Min and {Kassin}, Susan},
        title = "{The Imprint of Clump Formation at High Redshift. II. The Chemistry of the Bulge}",
      journal = {\apj},
         year = 2023,
        month = apr,
       volume = {946},
       number = {2},
          eid = {118},
        pages = {118},
          doi = {10.3847/1538-4357/acbb00},
archivePrefix = {arXiv},
       eprint = {2303.08265},
 primaryClass = {astro-ph.GA},
       adsurl = {https://ui.adsabs.harvard.edu/abs/2023ApJ...946..118D}
}

@ARTICLE{2014:Abreu,
       author = {{M{\'e}ndez-Abreu}, J. and {Debattista}, V.~P. and {Corsini}, E.~M. and {Aguerri}, J.~A.~L.},
        title = "{Secular- and merger-built bulges in barred galaxies}",
      journal = {\aap},
         year = 2014,
        month = dec,
       volume = {572},
          eid = {A25},
        pages = {A25},
          doi = {10.1051/0004-6361/201423955},
archivePrefix = {arXiv},
       eprint = {1409.2876},
 primaryClass = {astro-ph.GA},
       adsurl = {https://ui.adsabs.harvard.edu/abs/2014A&A...572A..25M}
}

@ARTICLE{2019:Lodders,
       author = {{Lodders}, Katharina},
        title = "{Solar Elemental Abundances}",
      journal = {arXiv e-prints},
         year = 2019,
        month = dec,
          eid = {arXiv:1912.00844},
        pages = {arXiv:1912.00844},
          doi = {10.48550/arXiv.1912.00844},
archivePrefix = {arXiv},
       eprint = {1912.00844},
 primaryClass = {astro-ph.SR},
       adsurl = {https://ui.adsabs.harvard.edu/abs/2019arXiv191200844L}
}

@ARTICLE{1962:Eggen,
       author = {{Eggen}, O.~J. and {Lynden-Bell}, D. and {Sandage}, A.~R.},
        title = "{Evidence from the motions of old stars that the Galaxy collapsed.}",
      journal = {\apj},
         year = 1962,
        month = nov,
       volume = {136},
        pages = {748},
          doi = {10.1086/147433},
       adsurl = {https://ui.adsabs.harvard.edu/abs/1962ApJ...136..748E}
}

@ARTICLE{2008:Hopkins,
       author = {{Hopkins}, Philip F. and {Hernquist}, Lars and {Cox}, Thomas J. and {Kere{\v{s}}}, Du{\v{s}}an},
        title = "{A Cosmological Framework for the Co-Evolution of Quasars, Supermassive Black Holes, and Elliptical Galaxies. I. Galaxy Mergers and Quasar Activity}",
      journal = {\apjs},
         year = 2008,
        month = apr,
       volume = {175},
       number = {2},
        pages = {356-389},
          doi = {10.1086/524362},
archivePrefix = {arXiv},
       eprint = {0706.1243},
 primaryClass = {astro-ph},
       adsurl = {https://ui.adsabs.harvard.edu/abs/2008ApJS..175..356H}
}

@ARTICLE{2004:Tremonti,
       author = {{Tremonti}, Christy A. and {Heckman}, Timothy M. and {Kauffmann}, Guinevere and {Brinchmann}, Jarle and {Charlot}, St{\'e}phane and {White}, Simon D.~M. and {Seibert}, Mark and {Peng}, Eric W. and {Schlegel}, David J. and {Uomoto}, Alan and {Fukugita}, Masataka and {Brinkmann}, Jon},
        title = "{The Origin of the Mass-Metallicity Relation: Insights from 53,000 Star-forming Galaxies in the Sloan Digital Sky Survey}",
      journal = {\apj},
         year = 2004,
        month = oct,
       volume = {613},
       number = {2},
        pages = {898-913},
          doi = {10.1086/423264},
archivePrefix = {arXiv},
       eprint = {astro-ph/0405537},
 primaryClass = {astro-ph},
       adsurl = {https://ui.adsabs.harvard.edu/abs/2004ApJ...613..898T}
}

@ARTICLE{2022:Zenocratti,
       author = {{Zenocratti}, L.~J. and {De Rossi}, M.~E. and {Theuns}, T. and {Lara-L{\'o}pez}, M.~A.},
        title = "{The origin of correlations between mass, metallicity, and morphology in galaxies from the EAGLE simulation}",
      journal = {\mnras},
         year = 2022,
        month = jun,
       volume = {512},
       number = {4},
        pages = {6164-6179},
          doi = {10.1093/mnras/stac906},
archivePrefix = {arXiv},
       eprint = {2203.16326},
 primaryClass = {astro-ph.GA},
       adsurl = {https://ui.adsabs.harvard.edu/abs/2022MNRAS.512.6164Z}
}

@ARTICLE{2005:Gallazzi,
       author = {{Gallazzi}, Anna and {Charlot}, St{\'e}phane and {Brinchmann}, Jarle and {White}, Simon D.~M. and {Tremonti}, Christy A.},
        title = "{The ages and metallicities of galaxies in the local universe}",
      journal = {\mnras},
         year = 2005,
        month = sep,
       volume = {362},
       number = {1},
        pages = {41-58},
          doi = {10.1111/j.1365-2966.2005.09321.x},
archivePrefix = {arXiv},
       eprint = {astro-ph/0506539},
 primaryClass = {astro-ph},
       adsurl = {https://ui.adsabs.harvard.edu/abs/2005MNRAS.362...41G}
}

@ARTICLE{2024:Stanton,
       author = {{Stanton}, T.~M. and {Cullen}, F. and {McLure}, R.~J. and {Shapley}, A.~E. and {Arellano-C{\'o}rdova}, K.~Z. and {Begley}, R. and {Amor{\'\i}n}, R. and {Barrufet}, L. and {Calabr{\`o}}, A. and {Carnall}, A.~C. and {Cirasuolo}, M. and {Dunlop}, J.~S. and {Donnan}, C.~T. and {Hamadouche}, M.~L. and {Liu}, F.~Y. and {McLeod}, D.~J. and {Pentericci}, L. and {Pozzetti}, L. and {Sanders}, R.~L. and {Scholte}, D. and {Topping}, M.~W.},
        title = "{The NIRVANDELS survey: the stellar and gas-phase mass-metallicity relations of star-forming galaxies at z = 3.5}",
      journal = {\mnras},
         year = 2024,
        month = aug,
       volume = {532},
       number = {3},
        pages = {3102-3119},
          doi = {10.1093/mnras/stae1705},
archivePrefix = {arXiv},
       eprint = {2405.00774},
 primaryClass = {astro-ph.GA},
       adsurl = {https://ui.adsabs.harvard.edu/abs/2024MNRAS.532.3102S}
}

@ARTICLE{2017:DeRossi,
       author = {{De Rossi}, Mar{\'\i}a Emilia and {Bower}, Richard G. and {Font}, Andreea S. and {Schaye}, Joop and {Theuns}, Tom},
        title = "{Galaxy metallicity scaling relations in the EAGLE simulations}",
      journal = {\mnras},
         year = 2017,
        month = dec,
       volume = {472},
       number = {3},
        pages = {3354-3377},
          doi = {10.1093/mnras/stx2158},
archivePrefix = {arXiv},
       eprint = {1704.00006},
 primaryClass = {astro-ph.GA},
       adsurl = {https://ui.adsabs.harvard.edu/abs/2017MNRAS.472.3354D}
}

@ARTICLE{2022:Gargiulo,
       author = {{Gargiulo}, Ignacio D. and {Monachesi}, Antonela and {G{\'o}mez}, Facundo A. and {Nelson}, Dylan and {Pillepich}, Annalisa and {Pakmor}, R{\"u}diger and {Grand}, R.~J.~J. and {Fragkoudi}, Francesca and {Hernquist}, Lars and {Lovell}, Mark and {Marinacci}, Federico},
        title = "{High and low S{\'e}rsic index bulges in Milky Way- and M31-like galaxies: origin and connection to the bar with TNG50}",
      journal = {\mnras},
         year = 2022,
        month = may,
       volume = {512},
       number = {2},
        pages = {2537-2555},
          doi = {10.1093/mnras/stac629},
archivePrefix = {arXiv},
       eprint = {2111.13712},
 primaryClass = {astro-ph.GA},
       adsurl = {https://ui.adsabs.harvard.edu/abs/2022MNRAS.512.2537G}
}

@ARTICLE{2011:Eliche,
       author = {{Eliche-Moral}, M.~C. and {Gonz{\'a}lez-Garc{\'\i}a}, A.~C. and {Balcells}, M. and {Aguerri}, J.~A.~L. and {Gallego}, J. and {Zamorano}, J. and {Prieto}, M.},
        title = "{A minor merger origin for stellar inner discs and rings in spiral galaxies}",
      journal = {\aap},
         year = 2011,
        month = sep,
       volume = {533},
          eid = {A104},
        pages = {A104},
          doi = {10.1051/0004-6361/201116509},
archivePrefix = {arXiv},
       eprint = {1105.5826},
 primaryClass = {astro-ph.CO},
       adsurl = {https://ui.adsabs.harvard.edu/abs/2011A&A...533A.104E}
}

@ARTICLE{2024:Angeloudi,
       author = {{Angeloudi}, Eirini and {Falc{\'o}n-Barroso}, Jes{\'u}s and {Huertas-Company}, Marc and {Boecker}, Alina and {Sarmiento}, Regina and {Eisert}, Lukas and {Pillepich}, Annalisa},
        title = "{Constraints on the in situ and ex situ stellar masses in nearby galaxies obtained with artificial intelligence}",
      journal = {Nature Astronomy},
         year = 2024,
        month = jul,
          doi = {10.1038/s41550-024-02327-3},
archivePrefix = {arXiv},
       eprint = {2407.00166},
 primaryClass = {astro-ph.GA},
       adsurl = {https://ui.adsabs.harvard.edu/abs/2024NatAs.tmp..197A}
}

@ARTICLE{1999:Noguchi,
       author = {{Noguchi}, Masafumi},
        title = "{Early Evolution of Disk Galaxies: Formation of Bulges in Clumpy Young Galactic Disks}",
      journal = {\apj},
         year = 1999,
        month = mar,
       volume = {514},
       number = {1},
        pages = {77-95},
          doi = {10.1086/306932},
archivePrefix = {arXiv},
       eprint = {astro-ph/9806355},
 primaryClass = {astro-ph},
       adsurl = {https://ui.adsabs.harvard.edu/abs/1999ApJ...514...77N}
}

@INPROCEEDINGS{2011:Bournaud,
       author = {{Bournaud}, F.},
        title = "{Star formation in galaxy interactions and mergers}",
    booktitle = {EAS Publications Series},
         year = 2011,
       editor = {{Charbonnel}, Corinne and {Montmerle}, Thierry},
       series = {EAS Publications Series},
       volume = {51},
        month = nov,
        pages = {107-131},
          doi = {10.1051/eas/1151008},
archivePrefix = {arXiv},
       eprint = {1106.1793},
 primaryClass = {astro-ph.CO},
       adsurl = {https://ui.adsabs.harvard.edu/abs/2011EAS....51..107B}
}

@ARTICLE{2019:Rosito,
       author = {{Rosito}, M.~S. and {Tissera}, P.~B. and {Pedrosa}, S.~E. and {Rosas-Guevara}, Y.},
        title = "{Assembly of spheroid-dominated galaxies in the EAGLE simulation}",
      journal = {\aap},
         year = 2019,
        month = sep,
       volume = {629},
          eid = {A37},
        pages = {A37},
          doi = {10.1051/0004-6361/201834720},
archivePrefix = {arXiv},
       eprint = {1811.11062},
 primaryClass = {astro-ph.GA},
       adsurl = {https://ui.adsabs.harvard.edu/abs/2019A&A...629A..37R}
}

@ARTICLE{2023:Cataldi,
       author = {{Cataldi}, P. and {Pedrosa}, S.~E. and {Tissera}, P.~B. and {Artale}, M.~C. and {Padilla}, N.~D. and {Dominguez-Tenreiro}, R. and {Bignone}, L. and {Gonzalez}, R. and {Pellizza}, L.~J.},
        title = "{Redshift evolution of the dark matter haloes shapes}",
      journal = {\mnras},
         year = 2023,
        month = aug,
       volume = {523},
       number = {2},
        pages = {1919-1932},
          doi = {10.1093/mnras/stad1601},
archivePrefix = {arXiv},
       eprint = {2302.08853},
 primaryClass = {astro-ph.GA},
       adsurl = {https://ui.adsabs.harvard.edu/abs/2023MNRAS.523.1919C}
}

@ARTICLE{2024:Casanueva,
       author = {{Casanueva-Villarreal}, C. and {Tissera}, P.~B. and {Padilla}, N. and {Liu}, B. and {Bromm}, V. and {Pedrosa}, S. and {Bignone}, L. and {Dominguez-Tenreiro}, R.},
        title = "{Impact of primordial black hole dark matter on gas properties at very high redshift: Semianalytical model}",
      journal = {\aap},
         year = 2024,
        month = aug,
       volume = {688},
          eid = {A183},
        pages = {A183},
          doi = {10.1051/0004-6361/202449650},
archivePrefix = {arXiv},
       eprint = {2405.02206},
 primaryClass = {astro-ph.GA},
       adsurl = {https://ui.adsabs.harvard.edu/abs/2024A&A...688A.183C}
}

@ARTICLE{2017:Fragkoudi,
       author = {{Fragkoudi}, F. and {Di Matteo}, P. and {Haywood}, M. and {G{\'o}mez}, A. and {Combes}, F. and {Katz}, D. and {Semelin}, B.},
        title = "{Bars and boxy/peanut bulges in thin and thick discs. I. Morphology and line-of-sight velocities of a fiducial model}",
      journal = {\aap},
         year = 2017,
        month = oct,
       volume = {606},
          eid = {A47},
        pages = {A47},
          doi = {10.1051/0004-6361/201630244},
archivePrefix = {arXiv},
       eprint = {1704.00734},
 primaryClass = {astro-ph.GA},
       adsurl = {https://ui.adsabs.harvard.edu/abs/2017A&A...606A..47F}
}

@ARTICLE{2014:Planck,
       author = {{Planck Collaboration} and {Ade}, P.~A.~R. and {Aghanim}, N. and {Armitage-Caplan}, C. and {Arnaud}, M. and {Ashdown}, M. and {Atrio-Barandela}, F. and {Aumont}, J. and {Baccigalupi}, C. and {Banday}, A.~J. and {Barreiro}, R.~B. and {Bartlett}, J.~G. and {Battaner}, E. and {Benabed}, K. and {Beno{\^\i}t}, A. and {Benoit-L{\'e}vy}, A. and {Bernard}, J. -P. and {Bersanelli}, M. and {Bielewicz}, P. and {Bobin}, J. and {Bock}, J.~J. and {Bonaldi}, A. and {Bond}, J.~R. and {Borrill}, J. and {Bouchet}, F.~R. and {Bridges}, M. and {Bucher}, M. and {Burigana}, C. and {Butler}, R.~C. and {Calabrese}, E. and {Cappellini}, B. and {Cardoso}, J. -F. and {Catalano}, A. and {Challinor}, A. and {Chamballu}, A. and {Chary}, R. -R. and {Chen}, X. and {Chiang}, H.~C. and {Chiang}, L. -Y. and {Christensen}, P.~R. and {Church}, S. and {Clements}, D.~L. and {Colombi}, S. and {Colombo}, L.~P.~L. and {Couchot}, F. and {Coulais}, A. and {Crill}, B.~P. and {Curto}, A. and {Cuttaia}, F. and {Danese}, L. and {Davies}, R.~D. and {Davis}, R.~J. and {de Bernardis}, P. and {de Rosa}, A. and {de Zotti}, G. and {Delabrouille}, J. and {Delouis}, J. -M. and {D{\'e}sert}, F. -X. and {Dickinson}, C. and {Diego}, J.~M. and {Dolag}, K. and {Dole}, H. and {Donzelli}, S. and {Dor{\'e}}, O. and {Douspis}, M. and {Dunkley}, J. and {Dupac}, X. and {Efstathiou}, G. and {Elsner}, F. and {En{\ss}lin}, T.~A. and {Eriksen}, H.~K. and {Finelli}, F. and {Forni}, O. and {Frailis}, M. and {Fraisse}, A.~A. and {Franceschi}, E. and {Gaier}, T.~C. and {Galeotta}, S. and {Galli}, S. and {Ganga}, K. and {Giard}, M. and {Giardino}, G. and {Giraud-H{\'e}raud}, Y. and {Gjerl{\o}w}, E. and {Gonz{\'a}lez-Nuevo}, J. and {G{\'o}rski}, K.~M. and {Gratton}, S. and {Gregorio}, A. and {Gruppuso}, A. and {Gudmundsson}, J.~E. and {Haissinski}, J. and {Hamann}, J. and {Hansen}, F.~K. and {Hanson}, D. and {Harrison}, D. and {Henrot-Versill{\'e}}, S. and {Hern{\'a}ndez-Monteagudo}, C. and {Herranz}, D. and {Hildebrandt}, S.~R. and {Hivon}, E. and {Hobson}, M. and {Holmes}, W.~A. and {Hornstrup}, A. and {Hou}, Z. and {Hovest}, W. and {Huffenberger}, K.~M. and {Jaffe}, A.~H. and {Jaffe}, T.~R. and {Jewell}, J. and {Jones}, W.~C. and {Juvela}, M. and {Keih{\"a}nen}, E. and {Keskitalo}, R. and {Kisner}, T.~S. and {Kneissl}, R. and {Knoche}, J. and {Knox}, L. and {Kunz}, M. and {Kurki-Suonio}, H. and {Lagache}, G. and {L{\"a}hteenm{\"a}ki}, A. and {Lamarre}, J. -M. and {Lasenby}, A. and {Lattanzi}, M. and {Laureijs}, R.~J. and {Lawrence}, C.~R. and {Leach}, S. and {Leahy}, J.~P. and {Leonardi}, R. and {Le{\'o}n-Tavares}, J. and {Lesgourgues}, J. and {Lewis}, A. and {Liguori}, M. and {Lilje}, P.~B. and {Linden-V{\o}rnle}, M. and {L{\'o}pez-Caniego}, M. and {Lubin}, P.~M. and {Mac{\'\i}as-P{\'e}rez}, J.~F. and {Maffei}, B. and {Maino}, D. and {Mandolesi}, N. and {Maris}, M. and {Marshall}, D.~J. and {Martin}, P.~G. and {Mart{\'\i}nez-Gonz{\'a}lez}, E. and {Masi}, S. and {Massardi}, M. and {Matarrese}, S. and {Matthai}, F. and {Mazzotta}, P. and {Meinhold}, P.~R. and {Melchiorri}, A. and {Melin}, J. -B. and {Mendes}, L. and {Menegoni}, E. and {Mennella}, A. and {Migliaccio}, M. and {Millea}, M. and {Mitra}, S. and {Miville-Desch{\^e}nes}, M. -A. and {Moneti}, A. and {Montier}, L. and {Morgante}, G. and {Mortlock}, D. and {Moss}, A. and {Munshi}, D. and {Murphy}, J.~A. and {Naselsky}, P. and {Nati}, F. and {Natoli}, P. and {Netterfield}, C.~B. and {N{\o}rgaard-Nielsen}, H.~U. and {Noviello}, F. and {Novikov}, D. and {Novikov}, I. and {O'Dwyer}, I.~J. and {Osborne}, S. and {Oxborrow}, C.~A. and {Paci}, F. and {Pagano}, L. and {Pajot}, F. and {Paladini}, R. and {Paoletti}, D. and {Partridge}, B. and {Pasian}, F. and {Patanchon}, G. and {Pearson}, D. and {Pearson}, T.~J. and {Peiris}, H.~V. and {Perdereau}, O. and {Perotto}, L. and {Perrotta}, F. and {Pettorino}, V. and {Piacentini}, F. and {Piat}, M. and {Pierpaoli}, E. and {Pietrobon}, D. and {Plaszczynski}, S. and {Platania}, P. and {Pointecouteau}, E. and {Polenta}, G. and {Ponthieu}, N. and {Popa}, L. and {Poutanen}, T. and {Pratt}, G.~W. and {Pr{\'e}zeau}, G. and {Prunet}, S. and {Puget}, J. -L. and {Rachen}, J.~P. and {Reach}, W.~T. and {Rebolo}, R. and {Reinecke}, M. and {Remazeilles}, M. and {Renault}, C. and {Ricciardi}, S. and {Riller}, T. and {Ristorcelli}, I. and {Rocha}, G. and {Rosset}, C. and {Roudier}, G. and {Rowan-Robinson}, M. and {Rubi{\~n}o-Mart{\'\i}n}, J.~A. and {Rusholme}, B. and {Sandri}, M. and {Santos}, D. and {Savelainen}, M. and {Savini}, G. and {Scott}, D. and {Seiffert}, M.~D. and {Shellard}, E.~P.~S. and {Spencer}, L.~D. and {Starck}, J. -L. and {Stolyarov}, V. and {Stompor}, R. and {Sudiwala}, R. and {Sunyaev}, R. and {Sureau}, F. and {Sutton}, D. and {Suur-Uski}, A. -S. and {Sygnet}, J. -F. and {Tauber}, J.~A. and {Tavagnacco}, D. and {Terenzi}, L. and {Toffolatti}, L. and {Tomasi}, M. and {Tristram}, M. and {Tucci}, M. and {Tuovinen}, J. and {T{\"u}rler}, M. and {Umana}, G. and {Valenziano}, L. and {Valiviita}, J. and {Van Tent}, B. and {Vielva}, P. and {Villa}, F. and {Vittorio}, N. and {Wade}, L.~A. and {Wandelt}, B.~D. and {Wehus}, I.~K. and {White}, M. and {White}, S.~D.~M. and {Wilkinson}, A. and {Yvon}, D. and {Zacchei}, A. and {Zonca}, A.},
        title = "{Planck 2013 results. XVI. Cosmological parameters}",
      journal = {\aap},
         year = 2014,
        month = nov,
       volume = {571},
          eid = {A16},
        pages = {A16},
          doi = {10.1051/0004-6361/201321591},
archivePrefix = {arXiv},
       eprint = {1303.5076},
 primaryClass = {astro-ph.CO},
       adsurl = {https://ui.adsabs.harvard.edu/abs/2014A&A...571A..16P}
}

@ARTICLE{2013:Cappellari,
       author = {{Cappellari}, Michele and {McDermid}, Richard M. and {Alatalo}, Katherine and {Blitz}, Leo and {Bois}, Maxime and {Bournaud}, Fr{\'e}d{\'e}ric and {Bureau}, M. and {Crocker}, Alison F. and {Davies}, Roger L. and {Davis}, Timothy A. and {de Zeeuw}, P.~T. and {Duc}, Pierre-Alain and {Emsellem}, Eric and {Khochfar}, Sadegh and {Krajnovi{\'c}}, Davor and {Kuntschner}, Harald and {Morganti}, Raffaella and {Naab}, Thorsten and {Oosterloo}, Tom and {Sarzi}, Marc and {Scott}, Nicholas and {Serra}, Paolo and {Weijmans}, Anne-Marie and {Young}, Lisa M.},
        title = "{The ATLAS$^{3D}$ project - XX. Mass-size and mass-{\ensuremath{\sigma}} distributions of early-type galaxies: bulge fraction drives kinematics, mass-to-light ratio, molecular gas fraction and stellar initial mass function}",
      journal = {\mnras},
         year = 2013,
        month = jul,
       volume = {432},
       number = {3},
        pages = {1862-1893},
          doi = {10.1093/mnras/stt644},
archivePrefix = {arXiv},
       eprint = {1208.3523},
 primaryClass = {astro-ph.CO},
       adsurl = {https://ui.adsabs.harvard.edu/abs/2013MNRAS.432.1862C}
}

@ARTICLE{2020:Bluck,
       author = {{Bluck}, Asa F.~L. and {Maiolino}, Roberto and {Piotrowska}, Joanna M. and {Trussler}, James and {Ellison}, Sara L. and {S{\'a}nchez}, Sebastian F. and {Thorp}, Mallory D. and {Teimoorinia}, Hossen and {Moreno}, Jorge and {Conselice}, Christopher J.},
        title = "{How do central and satellite galaxies quench? - Insights from spatially resolved spectroscopy in the MaNGA survey}",
      journal = {\mnras},
         year = 2020,
        month = nov,
       volume = {499},
       number = {1},
        pages = {230-268},
          doi = {10.1093/mnras/staa2806},
archivePrefix = {arXiv},
       eprint = {2009.05341},
 primaryClass = {astro-ph.GA},
       adsurl = {https://ui.adsabs.harvard.edu/abs/2020MNRAS.499..230B}
}

@INPROCEEDINGS{1999:Avila-Reese,
       author = {{Avila-Reese}, V. and {Firmani}, C.},
        title = "{On the Formation of Bulges and Elliptical Galaxies in the Cosmological Context}",
    booktitle = {Star Formation in Early Type Galaxies},
         year = 1999,
       editor = {{Carral}, P. and {Cepa}, J.},
       series = {Astronomical Society of the Pacific Conference Series},
       volume = {163},
        month = jan,
        pages = {243},
          doi = {10.48550/arXiv.astro-ph/9808165},
archivePrefix = {arXiv},
       eprint = {astro-ph/9808165},
 primaryClass = {astro-ph},
       adsurl = {https://ui.adsabs.harvard.edu/abs/1999ASPC..163..243A}
}

@ARTICLE{2013:Perez,
       author = {{Perez}, Josefa and {Valenzuela}, Octavio and {Tissera}, Patricia B. and {Michel-Dansac}, Leo},
        title = "{Clumpy disc and bulge formation}",
      journal = {\mnras},
         year = 2013,
        month = nov,
       volume = {436},
       number = {1},
        pages = {259-265},
          doi = {10.1093/mnras/stt1563},
archivePrefix = {arXiv},
       eprint = {1308.4396},
 primaryClass = {astro-ph.GA},
       adsurl = {https://ui.adsabs.harvard.edu/abs/2013MNRAS.436..259P}
}

@ARTICLE{2014:Zahid,
       author = {{Zahid}, H. Jabran and {Dima}, Gabriel I. and {Kudritzki}, Rolf-Peter and {Kewley}, Lisa J. and {Geller}, Margaret J. and {Hwang}, Ho Seong and {Silverman}, John D. and {Kashino}, Daichi},
        title = "{The Universal Relation of Galactic Chemical Evolution: The Origin of the Mass-Metallicity Relation}",
      journal = {\apj},
         year = 2014,
        month = aug,
       volume = {791},
       number = {2},
          eid = {130},
        pages = {130},
          doi = {10.1088/0004-637X/791/2/130},
archivePrefix = {arXiv},
       eprint = {1404.7526},
 primaryClass = {astro-ph.GA},
       adsurl = {https://ui.adsabs.harvard.edu/abs/2014ApJ...791..130Z}
}

@ARTICLE{2013:Kirby,
       author = {{Kirby}, Evan N. and {Cohen}, Judith G. and {Guhathakurta}, Puragra and {Cheng}, Lucy and {Bullock}, James S. and {Gallazzi}, Anna},
        title = "{The Universal Stellar Mass-Stellar Metallicity Relation for Dwarf Galaxies}",
      journal = {\apj},
         year = 2013,
        month = dec,
       volume = {779},
       number = {2},
          eid = {102},
        pages = {102},
          doi = {10.1088/0004-637X/779/2/102},
archivePrefix = {arXiv},
       eprint = {1310.0814},
 primaryClass = {astro-ph.GA},
       adsurl = {https://ui.adsabs.harvard.edu/abs/2013ApJ...779..102K}
}

@ARTICLE{2017:Johnston,
       author = {{Johnston}, Evelyn J. and {H{\"a}u{\ss}ler}, Boris and {Arag{\'o}n-Salamanca}, Alfonso and {Merrifield}, Michael R. and {Bamford}, Steven and {Bershady}, Matthew A. and {Bundy}, Kevin and {Drory}, Niv and {Fu}, Hai and {Law}, David and {Nitschelm}, Christian and {Thomas}, Daniel and {Roman Lopes}, Alexandre and {Wake}, David and {Yan}, Renbin},
        title = "{SDSS-IV MaNGA: bulge-disc decomposition of IFU data cubes (BUDDI)}",
      journal = {\mnras},
         year = 2017,
        month = feb,
       volume = {465},
       number = {2},
        pages = {2317-2341},
          doi = {10.1093/mnras/stw2823},
archivePrefix = {arXiv},
       eprint = {1611.00609},
 primaryClass = {astro-ph.GA},
       adsurl = {https://ui.adsabs.harvard.edu/abs/2017MNRAS.465.2317J}
}

@article{Yu:2023,
    author = {Yu, Sijie and Bullock, James S and Gurvich, Alexander B and Hafen, Zachary and Stern, Jonathan and Boylan-Kolchin, Michael and Faucher-Giguère, Claude-André and Wetzel, Andrew and Hopkins, Philip F and Moreno, Jorge},
    title = {Born this way: thin disc, thick disc, and isotropic spheroid formation in FIRE-2 Milky Way–mass galaxy simulations},
    journal = {Monthly Notices of the Royal Astronomical Society},
    volume = {523},
    number = {4},
    pages = {6220-6238},
    year = {2023},
    month = {06},
    issn = {0035-8711},
    doi = {10.1093/mnras/stad1806},
    url = {https://doi.org/10.1093/mnras/stad1806},
    eprint = {https://academic.oup.com/mnras/article-pdf/523/4/6220/50789794/stad1806.pdf},
}

@ARTICLE{1926:Hubble,
       author = {{Hubble}, E.~P.},
        title = "{Extragalactic nebulae.}",
      journal = {\apj},
         year = 1926,
        month = dec,
       volume = {64},
        pages = {321-369},
          doi = {10.1086/143018},
       adsurl = {https://ui.adsabs.harvard.edu/abs/1926ApJ....64..321H}
}

@ARTICLE{2025:Gonzalez,
       author = {{Gonzalez-Jara}, Jenny and {Tissera}, Patricia B. and {Monachesi}, Antonela and {Sillero}, Emanuel and {Pallero}, Diego and {Pedrosa}, Susana and {Tau}, Elisa A. and {Tapia-Contreras}, Brian and {Bignone}, Lucas},
        title = "{Unveiling the formation channels of stellar halos through their chemical fingerprints}",
      journal = {\aap},
         year = 2025,
        month = jan,
       volume = {693},
          eid = {A282},
        pages = {A282},
          doi = {10.1051/0004-6361/202452639},
archivePrefix = {arXiv},
       eprint = {2412.13483},
 primaryClass = {astro-ph.GA},
       adsurl = {https://ui.adsabs.harvard.edu/abs/2025A&A...693A.282G}
}

@ARTICLE{2021:Queiroz,
       author = {{Queiroz}, A.~B.~A. and {Chiappini}, C. and {Perez-Villegas}, A. and {Khalatyan}, A. and {Anders}, F. and {Barbuy}, B. and {Santiago}, B.~X. and {Steinmetz}, M. and {Cunha}, K. and {Schultheis}, M. and {Majewski}, S.~R. and {Minchev}, I. and {Minniti}, D. and {Beaton}, R.~L. and {Cohen}, R.~E. and {da Costa}, L.~N. and {Fern{\'a}ndez-Trincado}, J.~G. and {Garcia-Hern{\'a}ndez}, D.~A. and {Geisler}, D. and {Hasselquist}, S. and {Lane}, R.~R. and {Nitschelm}, C. and {Rojas-Arriagada}, A. and {Roman-Lopes}, A. and {Smith}, V. and {Zasowski}, G.},
        title = "{The Milky Way bar and bulge revealed by APOGEE and Gaia EDR3}",
      journal = {\aap},
         year = 2021,
        month = dec,
       volume = {656},
          eid = {A156},
        pages = {A156},
          doi = {10.1051/0004-6361/202039030},
archivePrefix = {arXiv},
       eprint = {2007.12915},
 primaryClass = {astro-ph.GA},
       adsurl = {https://ui.adsabs.harvard.edu/abs/2021A&A...656A.156Q}
}

@ARTICLE{2011:Guedes,
       author = {{Guedes}, Javiera and {Callegari}, Simone and {Madau}, Piero and {Mayer}, Lucio},
        title = "{Forming Realistic Late-type Spirals in a {\ensuremath{\Lambda}}CDM Universe: The Eris Simulation}",
      journal = {\apj},
         year = 2011,
        month = dec,
       volume = {742},
       number = {2},
          eid = {76},
        pages = {76},
          doi = {10.1088/0004-637X/742/2/76},
archivePrefix = {arXiv},
       eprint = {1103.6030},
 primaryClass = {astro-ph.CO},
       adsurl = {https://ui.adsabs.harvard.edu/abs/2011ApJ...742...76G}
}

@ARTICLE{1980:Tinsley,
       author = {{Tinsley}, B.~M.},
        title = "{Evolution of the Stars and Gas in Galaxies}",
      journal = {\fcp},
         year = 1980,
        month = jan,
       volume = {5},
        pages = {287-388},
          doi = {10.48550/arXiv.2203.02041},
archivePrefix = {arXiv},
       eprint = {2203.02041},
 primaryClass = {astro-ph.GA},
       adsurl = {https://ui.adsabs.harvard.edu/abs/1980FCPh....5..287T}
}

@BOOK{1997:Pagel,
       author = {{Pagel}, Bernard E.~J.},
        title = "{Nucleosynthesis and Chemical Evolution of Galaxies}",
         year = 1997,
       adsurl = {https://ui.adsabs.harvard.edu/abs/1997nceg.book.....P}
}

@ARTICLE{1997:Chiappini,
       author = {{Chiappini}, C. and {Matteucci}, F. and {Gratton}, R.},
        title = "{The Chemical Evolution of the Galaxy: The Two-Infall Model}",
      journal = {\apj},
         year = 1997,
        month = mar,
       volume = {477},
       number = {2},
        pages = {765-780},
          doi = {10.1086/303726},
archivePrefix = {arXiv},
       eprint = {astro-ph/9609199},
 primaryClass = {astro-ph},
       adsurl = {https://ui.adsabs.harvard.edu/abs/1997ApJ...477..765C}
}

@ARTICLE{2019:Maiolino,
       author = {{Maiolino}, R. and {Mannucci}, F.},
        title = "{De re metallica: the cosmic chemical evolution of galaxies}",
      journal = {\aapr},
         year = 2019,
        month = feb,
       volume = {27},
       number = {1},
          eid = {3},
        pages = {3},
          doi = {10.1007/s00159-018-0112-2},
archivePrefix = {arXiv},
       eprint = {1811.09642},
 primaryClass = {astro-ph.GA},
       adsurl = {https://ui.adsabs.harvard.edu/abs/2019A&ARv..27....3M}
}

@ARTICLE{2024:Lewis,
       author = {{Lewis}, Zach J. and {Andrews}, Brett H. and {Bezanson}, Rachel and {Maseda}, Michael and {Bell}, Eric F. and {Dav{\'e}}, Romeel and {D'Eugenio}, Francesco and {Franx}, Marijn and {Gallazzi}, Anna and {de Graaff}, Anna and {Kaushal}, Yasha and {Nersesian}, Angelos and {Newman}, Jeffrey A. and {van der Wel}, Arjen and {Wu}, Po-Feng},
        title = "{The Gas-phase Mass{\textendash}Metallicity Relation for Massive Galaxies at z {\ensuremath{\sim}} 0.7 with the LEGA-C Survey}",
      journal = {\apj},
         year = 2024,
        month = mar,
       volume = {964},
       number = {1},
          eid = {59},
        pages = {59},
          doi = {10.3847/1538-4357/ad250c},
archivePrefix = {arXiv},
       eprint = {2304.12343},
 primaryClass = {astro-ph.GA},
       adsurl = {https://ui.adsabs.harvard.edu/abs/2024ApJ...964...59L}
}

@ARTICLE{2024:Cheng,
       author = {{Cheng}, Yingjie and {Giavalisco}, Mauro and {Simons}, Raymond C. and {Ji}, Zhiyuan and {Stroupe}, Darren and {Cleri}, Nikko J.},
        title = "{Exploring the Gas-phase Metallicity Gradients of Star-forming Galaxies at Cosmic Noon}",
      journal = {\apj},
         year = 2024,
        month = mar,
       volume = {964},
       number = {1},
          eid = {94},
        pages = {94},
          doi = {10.3847/1538-4357/ad234a},
archivePrefix = {arXiv},
       eprint = {2401.12319},
 primaryClass = {astro-ph.GA},
       adsurl = {https://ui.adsabs.harvard.edu/abs/2024ApJ...964...94C}
}

@ARTICLE{2023:Baker,
       author = {{Baker}, William M. and {Maiolino}, Roberto and {Belfiore}, Francesco and {Curti}, Mirko and {Bluck}, Asa F.~L. and {Lin}, Lihwai and {Ellison}, Sara L. and {Thorp}, Mallory and {Pan}, Hsi-An},
        title = "{The metallicity's fundamental dependence on both local and global galactic quantities}",
      journal = {\mnras},
         year = 2023,
        month = feb,
       volume = {519},
       number = {1},
        pages = {1149-1170},
          doi = {10.1093/mnras/stac3594},
archivePrefix = {arXiv},
       eprint = {2210.03755},
 primaryClass = {astro-ph.GA},
       adsurl = {https://ui.adsabs.harvard.edu/abs/2023MNRAS.519.1149B}
}

@ARTICLE{2025:Tissera,
       author = {{Tissera}, Patricia B. and {Bignone}, Lucas and {Gonzalez-Jara}, Jenny and {Mu{\~n}oz-Escobar}, Ignacio and {Cataldi}, Pedro and {Miranda}, Valentina P. and {Barrientos-Acevedo}, Daniela and {Tapia-Contreras}, Brian and {Pedrosa}, Susana and {Padilla}, Nelson and {Dominguez-Tenreiro}, Rosa and {Casanueva-Villarreal}, Catalina and {Sillero}, Emanuel and {Silva-Mella}, Benjamin and {Shailesh}, Isha and {Jara-Ferreira}, Francisco},
        title = "{The CIELO project: The chemo-dynamical properties of galaxies and the cosmic web}",
      journal = {\aap},
         year = 2025,
        month = may,
       volume = {697},
          eid = {A134},
        pages = {A134},
          doi = {10.1051/0004-6361/202453348},
archivePrefix = {arXiv},
       eprint = {2501.05978},
 primaryClass = {astro-ph.GA},
       adsurl = {https://ui.adsabs.harvard.edu/abs/2025A&A...697A.134T}
}

\begin{appendix} 
\section{Effects of the bulge/halo decomposition}\label{sec:ap1}

\begin{figure}[h]
   \centering
        \includegraphics[width = 0.45\textwidth]{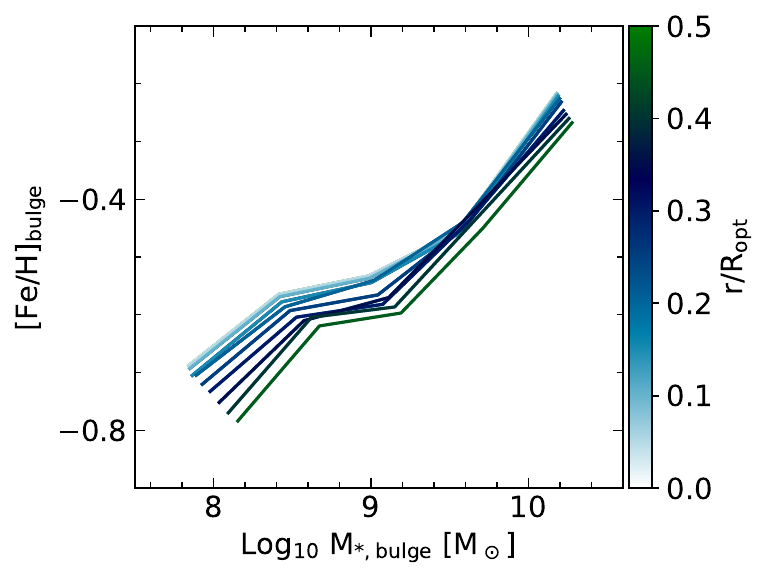}
         \caption{\sMZR~of the bulge including the halo up to different radii. The color-code of each line represents the corresponding radius.}.\label{fig:mzrwhalo}
   \end{figure}

The decomposition between bulge and disk relies on the binding energies at 0.5\ropt, as defined in Sec. \ref{sec:decomposition}, and the circularity. This allow us to differentiate between inner-halo, less graviationally bounded, and bulge, more bounded. Nevertheless, to better quantify the effects of considering the inner-halo as bulge taking into account that observers usually take  a fixed radius,  Fig.~\ref{fig:mzrwhalo} shows the different \sMZR~of the bulges, when considering all stars within a given radius as bulge (without separating components). As shown in Fig.~\ref{fig:mzrwhalo}, considering the inner halo  produces a decrease of  the overall metallicity of the bulge. This is expected since the halo has shown to be formed by mainly ex-situ stars that have lower metallicities \citep{2025:Gonzalez}. However, except from the low-mass end, there are no significant changes in the shape.

\section{\sMZR~and galaxy stellar mass}\label{sec:ap2}

\begin{figure}[h]
   \centering
        \includegraphics[width = 0.45\textwidth]{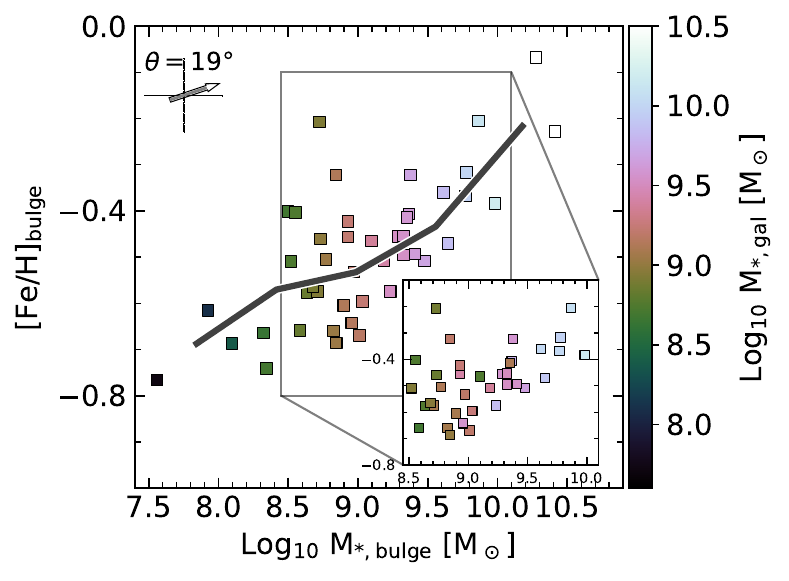}
         \caption{\sMZR~for the simulated bulges color-coded by the galaxy stellar mass. Similarly to Fig. \ref{fig:mzr_facc}, LOESS2D and the PCC analysis are included to quantify the trends with the third dependence. The inset figure shows the distribution without LOESS2D for the mass range of interest as comparison.}
         \label{fig:mzr_galmass}
   \end{figure}

The dependence with the fraction of stars formed ex-situ might also be affected by the stellar mass of the galaxy itself. To test if the dispersion at a given bulge mass also correlates with the galaxy mass, Fig.~\ref{fig:mzr_galmass} shows the \sMZR~color-coded by the galaxy mass. As expected larger bulges are found in larger galaxies. 
Nevertheless, the PCC analysis show an small positive angle, highlighting that the dependence on the stellar mass of a galaxy is small and the correlation with \fout~is stronger at a given bulge mass.

\section{Properties of the major contributors satellites.}\label{sec:ap3}

\begin{figure}[h]
   \centering
        \includegraphics[width = 0.45\textwidth]{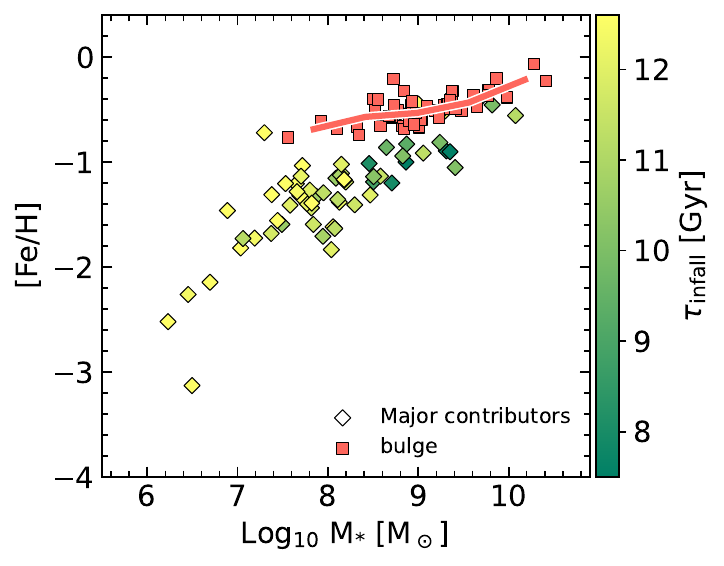}
         \caption{\sMZR~for the simulated bulges (orange squares) and the satellites (diamonds) that contributes to the bulge accreted stellar populations. The color-code correspond to the infall times of the contributor.}
         \label{fig:mzr_contributors}
   \end{figure}

The ISM from which stellar populations form is expected to determine the chemical enrichment that they have. In that sense, the accreted stars are born from gas that is expected to be from lower mass satellites or less enriched. Figure ~\ref{fig:mzr_contributors} shows the \sMZR~of the 3 major contributors for the bulges in the sample, at their time of infall. Additionally, the bulge \sMZR~of the central galaxies at $z = 0$ is included (orange squares). This only considers those satellites that had more than 100 stellar particles to avoid numerical noise. As expected from Fig.~\ref{fig:MZR_origin} they tend to have lower masses, and to be more metal poor. Furthermore, as we move to lower stellar masses, satellites tend to infall at earlier times of the simulation . The median infall time of the satellites is $11.6^{12.6}_{9.6}$ Gyr in lookback time (lower and upper numbers represent the 16$^\mathrm{th}$ and
84$^\mathrm{th}$ percentiles). These times are high, indicating that the stellar population formed early. This is consistent with their lower metallicities in comparison to the global populations of the bulge at $z = 0$.

\end{appendix}
\end{document}